\documentclass[11pt]{article}
\usepackage{aas_macros,amsmath,amssymb,comment,cite,esint,graphicx,mathtools}
\usepackage[margin=.8in,letterpaper]{geometry}
\usepackage[colorlinks=true]{hyperref}
\usepackage[affil-it]{authblk}
\usepackage{subcaption}
\usepackage[utf8]{inputenc}
\usepackage{mathrsfs}
\usepackage{appendix}
\usepackage{amssymb}
\usepackage{float}
\usepackage{color}
\usepackage{cite}
\usepackage{hyperref}
\hypersetup{pageanchor=false}
\usepackage{indentfirst}
\usepackage{url}
\usepackage{float}
\usepackage{caption}
\usepackage[numbers,square,comma,sort&compress,merge]{natbib}
\usepackage{esint}
\usepackage{overpic}
\usepackage{graphicx}
\usepackage{epsf,amsmath,bbold,amsfonts,stmaryrd}
\usepackage{textcomp}
\usepackage{ulem}
\usepackage{tikz}
\usepackage{booktabs}

\numberwithin{equation}{section}
\let\originalleft\left
\let\originalright\right
\renewcommand{\left}{\mathopen{}\mathclose\bgroup\originalleft}
\renewcommand{\right}{\aftergroup\egroup\originalright}
\mathcode`\*="8000
{\catcode`\*=\active\gdef*{\mathclose{}\,\mathopen{}}}

\newcommand{\bea}{\setlength\arraycolsep{2pt} \begin{eqnarray}}
\newcommand{\eea}{\end{eqnarray}}

\def\bag{\begin{aligned}}
\def\eag{\end{aligned}}
\def\bea{\begin{eqnarray}}
\def\eea{\end{eqnarray}}
\def\ba{\begin{array}}
\def\ea{\end{array}}

\def\bc{\begin{center}}
\def\ec{\end{center}}

\begin{document}
\title{Anomalous Increase in the Shadow Area of Rotating Ho\v{r}ava Black Holes}
	
\author{Zhi-Wu Zhou$^{1}$, Xin-Yang Wang$^{2, 4}$, Minyong Guo$^{3, 4}$, Wen-Biao Liu$^{2, 3}$}
\date{}
	
\maketitle
\vspace{-10mm}

\begin{center}
{\it
$^1$ School of Mathematics, Physics and Artificial Intelligence, China University of Petroleum-Beijing at Karamay, Karamay 834000, China\\\vspace{4mm}

$^2$ Faculty of Arts and Sciences, Beijing Normal University, Zhuhai 519087, China\\\vspace{4mm}

$^3$ School of Physics and Astronomy, Beijing Normal University, Beijing 100875, China\\\vspace{4mm}

$^4$ Key Laboratory of Multiscale Spin Physics (Beijing Normal University), Ministry of Education, Beijing 100875, China\\\vspace{4mm}
}
\end{center}

\vspace{8mm}

\begin{abstract}
    The black hole shadow area provides a natural global measure of the overall shadow size, although analytical calculations remain challenging in rotating spacetimes. An approximate separation method for the Hamilton-Jacobi equation is introduced to analytically construct the shadows of rotating Ho\v{r}ava black holes for observers at both spatial infinity and finite radial distances, and an asymptotic relationship between the projection schemes adopted at finite radial distances and spatial infinity is established. The shadow area integral is evaluated using Green's theorem, revealing an anomalous increase in the shadow area as the spin parameter $a$ increases at large observer inclination angles and high values of the Lorentz violation (LV) factor. At spatial infinity, a power series expansion in $a$ reveals that the terms involving the LV factor cancel in the second-order coefficient, while the anomalous increase in the shadow area is primarily governed by the fourth-order term. At finite radial distances, this cancellation no longer occurs, and the second-order term becomes the dominant contribution to the anomalous increase. The dependence of the second-order coefficient on the LV factor is further shown to originate exclusively from photon dynamics rather than the image plane projection.
\end{abstract}

\vfill{\footnotesize \flushleft $\ast$ Corresponding author: xinyangwang@bnu.edu.cn
}

\maketitle
\newpage
\baselineskip 18pt

\section{Introduction}\label{intro}

Black hole shadows provide a direct probe of photon propagation in the strong gravitational fields surrounding black holes and have become an important observational tool for testing gravitational theories on event horizon scales \cite{EventHorizonTelescope:2019dse, EventHorizonTelescope:2022xqj, Perlick:2021aok}. Within the geometric optics approximation and under appropriate background illumination conditions, the shadow boundary can be characterized by the critical curve on the observer's image plane \cite{Grenzebach:2014fha, Gralla:2019xty, Bronzwaer:2021lzo, Peng:2020wun}. The critical curve is associated with photon trajectories that asymptotically approach unstable photon orbits near the black hole, thereby connecting the properties of these orbits to the observable shadow contour \cite{Guo:2020zmf}. Consequently, the size and shape of the black hole shadow can provide information about the underlying spacetime geometry \cite{Hioki:2009na, Guo:2018kis}. To quantify these geometric properties, various observables have been introduced, including the horizontal and vertical diameters, displacement, deviation from circularity, distortion parameter, and mean radius \cite{Hioki:2009na, Zhong:2021mty, Perlick:2021aok, Hou:2021okc}. In contrast to observables defined primarily by a limited number of characteristic points or an effective radius, the area enclosed by the critical curve provides a natural global observable \cite{Cunha:2019ikd, Kumar:2020owy, Zhang:2023bzv}. By incorporating contributions from the entire critical curve, the shadow area provides a direct and geometrically well defined measure of the overall shadow size \cite{Gralla:2019xty, Perlick:2021aok}.

For a Schwarzschild black hole, spherical symmetry guarantees a circular critical curve whose radius can be obtained analytically, allowing the shadow area to be determined directly \cite{Synge:1966okc, Perlick:2021aok}. For rotating black holes, however, a general closed form expression for the shadow area is not readily available for arbitrary spin parameters and observer inclination angles, even in Kerr spacetime, although the unstable spherical photon orbits and the corresponding critical curve admit analytical parametric representations \cite{Teo:2003ltt, deVries:1999tiy, Grenzebach:2014fha, Perlick:2021aok}. Consequently, the area integral can generally be evaluated analytically only under specific conditions \cite{Feng:2019zzn}. Alternatively, black hole shadows can be constructed using numerical backward ray tracing, with the shadow area estimated by counting pixels enclosed by the critical curve \cite{Cunha:2016bpi, Younsi:2016azx}. Although numerical methods can reproduce the dependence of the shadow area on black hole parameters, the accuracy of these methods depends on numerical resolution and boundary extraction, while explicit functional relationships remain difficult to obtain. These limitations motivate the development of analytical or semi-analytical methods applicable to arbitrary observer inclination angles and spin parameters, enabling a systematic investigation of how the shadow area depends on spacetime parameters.

Modified theories of gravity provide a framework for investigating quantum gravity and possible deviations from general relativity (GR) in strong gravitational fields \cite{Bambi:2015kza}. Black hole shadows offer a means of testing the Kerr hypothesis and constraining deviations from Kerr geometry through observations \cite{Johannsen:2010ru, Mizuno:2018lxz, EventHorizonTelescope:2020qrl}. Among these theories, Ho\v{r}ava gravity has attracted considerable attention as a power counting renormalizable approach to quantum gravity \cite{Horava:2009uw, Sotiriou:2010wn, Wang:2017brl}. The theory introduces anisotropic Lifshitz scaling between space and time, allowing higher order spatial derivatives to improve ultraviolet behavior while restricting time derivatives to second order to avoid ghost instabilities \cite{Horava:2009uw, Blas:2009qj}. Perturbative renormalizability has been established for the projectable formulation, although the corresponding question remains more subtle in the more general formulation \cite{Barvinsky:2015kil}. This construction introduces a preferred foliation of spacetime, restricting the diffeomorphism invariance of GR to foliation preserving diffeomorphisms and breaking local Lorentz invariance \cite{Sotiriou:2010wn, Wang:2017brl, Barausse:2011pu}. At low energy scales, higher order spatial derivative operators are suppressed, while the preferred foliation and Lorentz violating couplings remain, with GR recovered only in the appropriate limit of these couplings \cite{Jacobson:2010mx}. Ho\v{r}ava black holes therefore provide a useful framework for investigating the effects of Lorentz violation (LV) on photon propagation and identifying possible observational signatures in strong gravitational fields.

In the low energy regime of four dimensional Ho\v{r}ava gravity, Devecio\u{g}lu and Park obtained an exact rotating black hole solution of Kerr type, referred to as the rotating Ho\v{r}ava black hole \cite{Devecioglu:2024Rotating}. The dimensionless combination $\zeta \equiv \kappa \xi$, hereafter termed the LV factor, governs deviations from Kerr geometry, and the solution reduces to the Kerr geometry when $\zeta = 1$. The influence of the LV factor on photon propagation in strong gravitational fields is reflected in the rotating Ho\v{r}ava black hole shadow, with the degree of deviation from Kerr geometry depending on the value of $\zeta$. Theoretical shadow predictions can further be compared with observational images to examine the influence of the LV factor on photon propagation in strong gravitational fields and constrain possible departures from the Kerr hypothesis for astrophysical black holes on event horizon scales \cite{Medeiros:2020NonKerr, Psaltis:2021Metric}.

To explore these potential observational signatures, several studies have investigated the shadow and imaging properties of black holes in Ho\v{r}ava gravity. The shadow of a rotating Ho\v{r}ava-Lifshitz black hole was examined in Ref. \cite{Atamurotov:2013dpa}, showing that gravitational parameters can increase the shadow size while reducing distortion relative to the Kerr case. Subsequently, the shadow of the rotating Ho\v{r}ava black hole was constructed using a numerical method based on backward ray tracing, and Event Horizon Telescope observations of M87* and Sgr A* were employed to investigate constraints on the LV factor \cite{Liu:2025lwj}. Furthermore, thin disk images and polarization signatures were investigated, revealing the influence of the LV factor on the inner shadow, brightness asymmetry, critical curve, and polarization distribution \cite{Zhao:2025nwi}. However, previous studies of the rotating Ho\v{r}ava black hole have primarily relied on numerical methods, while analytical investigations of the shadow area remain limited. Therefore, an analytical method is developed to investigate the dependence of the shadow area on the LV factor.

The remainder of this paper is organized as follows. Sec. \ref{section2} introduces the rotating Ho\v{r}ava black hole metric and the approximate separation of the Hamilton-Jacobi equation governing photon motion. In Sec. \ref{section3}, black hole shadows are constructed for observers at spatial infinity and finite radial distances using approximate analytical and numerical methods, and the relationship between the corresponding shadow images is established. Sec. \ref{section4} investigates the dependence of the shadow area on the observer inclination angle, black hole spin parameter, and LV factor for both observer configurations. The paper ends with conclusions in Sec. \ref{section5}.

In this study, the fundamental constants $c$ and $G$ are set to unity for simplicity, and the spacetime metric signature is adopted as $(-,+,+,+)$.

\section{Rotating Ho\v{r}ava spacetime and photon dynamics}\label{section2}

In Boyer Lindquist (BL) coordinates, the line element of the rotating Ho\v{r}ava spacetime is
\begin{equation}\label{rotatinghoravametric}
  \begin{split}
    ds^2 = & \left[- \frac{\Delta_r - a^2 \sin^2 \theta}{\rho^2} + \frac{(\zeta - 1)(2 m r)^2 a^2 \sin^2 \theta}{\rho^2 \Sigma^2} \right] d t^2 + \frac{\rho^2}{\Delta_r} dr^2 + \rho^2 d \theta^2\\
    &+ \frac{\Sigma^2 \sin^2 \theta}{\rho^2} d \phi^2 - \frac{4 a m r \sqrt{\zeta} \sin^2 \theta}{\rho^2} dt d \phi\,,
  \end{split}
\end{equation}
where
\begin{equation}
  \begin{split}
    \Delta_r & = r^2 + a^2 - 2 m r\,,\\
    \rho^2 & = r^2 + a^2 \cos^2 \theta\,,\\
    \Sigma^2 & = (r^2 + a^2) \rho^2 + 2 m r a^2 \sin^2 \theta\,.
  \end{split}
\end{equation}
The parameter range $1 \leq \zeta \leq4$ is adopted by restricting the analysis to the positive branch of the LV factor \cite{Zhao:2025nwi}. The line element reduces to the Kerr solution for $\zeta = 1$ and to the Schwarzschild solution for $a = 0$. For simplicity and without loss of generality, the parameter $m$ appearing in the line element is set to $m = 1$ throughout the subsequent discussion.

The construction of rotating Ho\v{r}ava black hole shadows requires determining photon trajectories governed by null geodesics in the rotating Ho\v{r}ava spacetime. The Kerr spacetime is of Petrov type D, and the Hamilton-Jacobi equation governing photon motion is separable, allowing the Kerr black hole shadow to be constructed analytically. In contrast, the rotating Ho\v{r}ava spacetime is generally of Petrov type I, and the corresponding Hamilton-Jacobi equation is inseparable, which poses challenges for the direct analytical construction of the black hole shadow. Nevertheless, the rotating Ho\v{r}ava spacetime reduces to the Kerr geometry when the LV factor $\zeta = 1$, indicating that the rotating Ho\v{r}ava geometry can be regarded as a deformation of the Kerr spacetime governed by the LV factor. This close relationship motivates an investigation of whether an analytical approach analogous to that employed for the Kerr black hole can be developed to construct the rotating Ho\v{r}ava black hole shadow.

The Hamilton-Jacobi equation governing the motion of a particle can be written as
\begin{equation}\label{hjequation}
  \frac{\partial S}{\partial \sigma} + H = 0\,,
\end{equation}
where $S$ is the action of the particle, $\sigma$ is an appropriately chosen affine parameter, and $H$ is the Hamiltonian of the particle. For geodesic motion in curved spacetime, the Hamiltonian can be expressed as
\begin{equation}\label{definitionhamiltonian}
    H = \frac{1}{2} g^{a b} \frac{\partial S}{\partial x^a} \frac{\partial S}{\partial x^b} = \frac{1}{2} g^{a b} p_a p_b = - \frac{1}{2} \mu^2\,.
\end{equation}
In Eq. (\ref{definitionhamiltonian}), $p_a$ denotes the covariant component of the four momentum of the particle, defined in terms of the action $S$ as
\begin{equation}
    p_a = \frac{\partial S}{\partial x^a}\,,
\end{equation}
and $\mu$ denotes the mass of the particle. For a timelike particle, $\mu$ is nonzero, whereas for a photon, $\mu$ vanishes. The Hamilton-Jacobi equation governing the particle in this spacetime is initially assumed to be separable, and the action $S$ can be written as
\begin{equation}\label{actionofparticle}
    \begin{split}
        S = \frac{1}{2} \mu^2 \sigma - E t + L \phi + S_r (r) + S_\theta (\theta)\,.
    \end{split}
\end{equation}
where $E = -p_a (\partial_t)^a$ and $L = p_a (\partial_\phi)^a$ denote the conserved energy and angular momentum of the particle associated with the spacetime Killing vectors $(\partial_t)^a$ and $(\partial_\phi)^a$, respectively. Substituting Eqs. (\ref{rotatinghoravametric}) and (\ref{actionofparticle}) into Eq. (\ref{hjequation}), the Hamilton-Jacobi equation governing particle motion in the rotating Ho\v{r}ava spacetime can be obtained as
\begin{equation}\label{resultinghjequation}
    \begin{split}
    & \left \{\Delta_r \left(\frac{dS_r}{dr} \right)^2 - \frac{\left[(r^2 + a^2) E - a L \right]^2}{\Delta_r} + \mu^2 r^2 + \frac{4 a r (\sqrt{\zeta} - 1)}{\Delta_r} E L \right \}\\
    & + \left \{ \left(\frac{dS_\theta}{d\theta} \right)^2 +\frac{\left(a \sin^2 \theta E - L \right)^2}{\sin^2 \theta} + a^2 \mu^2 \cos^2 \theta \right \} + T(r,\, \theta) = 0\,,
  \end{split}
\end{equation}
where the function $T (r,\, \theta)$ is given by
\begin{equation}\label{functiontrtheta}
  T (r,\, \theta) = \frac{4 (1 - \zeta) a^2 r^2}{\Delta_r \Sigma^2} L^2.
\end{equation}

In Eq. (\ref{resultinghjequation}), the terms enclosed in the first pair of braces depend exclusively on the radial coordinate $r$, whereas those in the second pair depend solely on the angular coordinate $\theta$. However, the function $T(r,\, \theta)$ depends on both coordinates, introducing a coupling between the radial and angular parts of the Hamilton-Jacobi equation. Consequently, this function prevents the exact separation of variables and poses challenges for the analytical description of geodesic motion in the rotating Ho\v{r}ava spacetime. To investigate the feasibility of approximate variable separation, the radial and angular dependence of $T(r,\, \theta)$ is analyzed to determine whether the function can be incorporated into the corresponding part of the Hamilton-Jacobi equation. As shown in Eq. (\ref{functiontrtheta}), the conserved angular momentum $L$ appears only as an overall factor and therefore does not affect the coordinate dependence of $T(r,\, \theta)$. For nonzero $L$, the function can be normalized by $L^2$ to define $\widetilde{T}(r,\, \theta) = T(r,\, \theta)/L^2$. The dependence of $\widetilde{T}(r,\, \theta)$ on both coordinates can be investigated by examining the two dimensional distribution in the $(r,\, \theta)$ plane. However, a direct visualization of this distribution obscures the distinction between the radial and angular dependence, particularly at small radial coordinates. To overcome this limitation, a logarithmic transformation is applied by considering $\log_{10}|\widetilde{T}(r,\, \theta)|$, allowing the coordinate dependence to be visualized more clearly.

\begin{figure}[htbp]
    \centering
    \begin{subfigure}[b]{0.48\textwidth}
        \includegraphics[width=\textwidth]{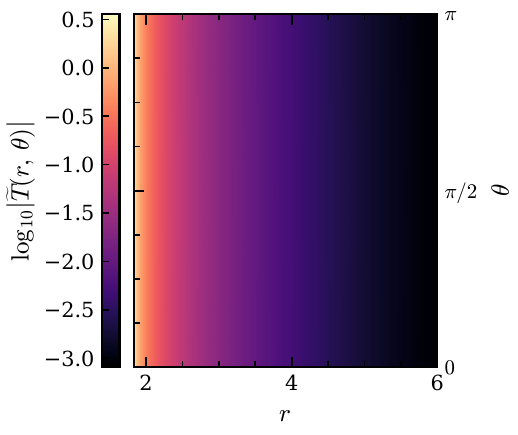}
        \caption{}
    \end{subfigure}
    \begin{subfigure}[b]{0.48\textwidth}
        \includegraphics[width=\textwidth]{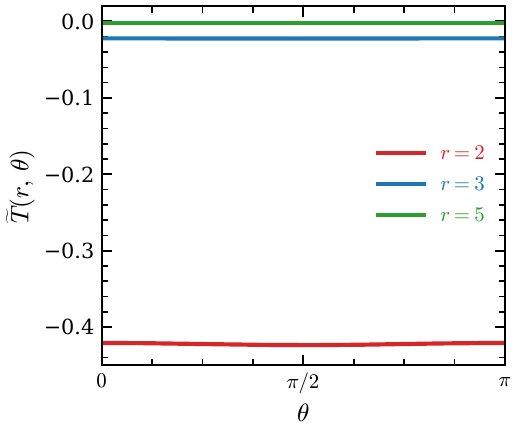}
        \caption{}
    \end{subfigure}
    \caption{The dependence of the function $\widetilde{T}(r,\, \theta)$ on the radial coordinate $r$ and the angular coordinate $\theta$ is illustrated for the parameter values $a = 0.6$ and $\zeta = 1.5$. Panel (a) presents the distribution of $\log_{10}|\widetilde{T}(r,\, \theta)|$ in the $(r,\, \theta)$ plane, whereas panel (b) shows the angular variation of $\widetilde{T}(r,\, \theta)$ at the fixed radial coordinates $r = 2$, $3$, and $5$.}
    \label{fig1}
\end{figure}
Fig. \ref{fig1} (a) presents the distribution of $\log_{10}|\widetilde{T}(r,\, \theta)|$ in the $(r,\, \theta)$ plane for the spin parameter $a = 0.6$ and LV factor $\zeta = 1.5$. As shown in the figure, the function exhibits significant variation with the radial coordinate $r$ at fixed $\theta$, whereas the variation with the angular coordinate $\theta$ at fixed $r$ is negligible. This indicates that $\widetilde{T}(r,\, \theta)$ depends predominantly on the radial coordinate $r$, with only a weak dependence on $\theta$. Accordingly, the radial dependence dominates $T(r,\, \theta)$. Fig. \ref{fig1} (b) further illustrates the angular dependence of $\widetilde{T}(r,\, \theta)$ at fixed radial coordinates. For $r = 2$, $3$, and $5$, the function remains nearly constant over the considered range of $\theta$, confirming that the angular dependence is sufficiently weak to be neglected.

Although the radial dependence of $\widetilde{T}(r,\, \theta)$ dominates over the angular dependence for $a = 0.6$ and $\zeta = 1.5$, the function $\widetilde{T}(r,\, \theta)$ also depends on the spin parameter $a$ and LV factor $\zeta$. Since $\zeta$ enters $\widetilde{T}(r,\, \theta)$ only as an overall multiplicative factor, variations in the LV factor do not affect the relative radial and angular dependence of the function. By contrast, the nonlinear dependence of $\widetilde{T}(r,\, \theta)$ on $a$ may affect the relative radial and angular dependence, requiring further examination of whether the angular dependence remains negligible for different values of $a$. To characterize the influence of $a$ on the relative radial and angular dependence of $\widetilde{T}(r,\, \theta)$ independently of the LV factor, a relative variation function $\mathcal{M}(r,\, \theta;\, a)$ can be defined as
\begin{equation}\label{functionmrthetaa}
  \mathcal{M} (r\,, \theta\,; a) \equiv \frac{\widetilde{T}(r\,, \theta)}{\widetilde{T}(r\,, 0)} - 1 = \frac{a^2 \Delta_r \sin^2 \theta}{(r^2 + a^2)^2 - a^2 \Delta_r \sin^2 \theta}.
\end{equation}
The function $\mathcal{M}(r,\, \theta;\, a)$ quantifies the relative variation of $\widetilde{T}(r,\, \theta)$ with respect to the reference value at $\theta = 0$ for a fixed radial coordinate $r$. The dependence of $\mathcal{M}(r,\, \theta;\, a)$ on $r$, $\theta$, and $a$ allows a systematic assessment of whether the angular dependence of $\widetilde{T}(r,\, \theta)$ remains negligible for different values of the spin parameter $a$. The function $\mathcal{M}(r,\, \theta;\, a)$ can be shown to increase monotonically with $\theta$ over the interval $\theta \in [0,\, \pi / 2]$ for fixed values of $a$ and $r$. This monotonicity implies that $\mathcal{M}(r,\, \pi / 2;\, a)$ characterizes the maximum relative angular variation at each radial position, providing a quantitative criterion for assessing whether the angular dependence of $\widetilde{T}(r,\, \theta)$ can be neglected. To evaluate the validity of this approximation over the radial domain, the maximum value $\mathcal{M}_{\max}$ and the corresponding radial position $r_{\max}$ are determined numerically for different values of $a$ using Eq. (\ref{functionmrthetaa}). The results are summarized in Table \ref{table1}.
\begin{table}[htbp]
\centering
\caption{Maximum relative variation $\mathcal{M}_{\max}$ and corresponding radial position $r_{\max}$ for different values of the spin parameter $a$.}
\label{table1}
\begin{tabular}{ccccc}
\toprule
$a$ & $0.2$ & $0.6$ & $0.9$ & $0.998$ \\
\midrule
$r_{\text{max}}$ & $2.98209$ & $2.82766$ & $2.56023$ & $2.41864$ \\
$\mathcal{M}_{\text{max}}$ & $0.1490\%$ & $1.4120\%$ & $3.4678\%$ & $4.4590\%$ \\
\bottomrule
\end{tabular}
\end{table}
As presented in Table \ref{table1}, $\mathcal{M}_{\max}$ exhibits a monotonic increase with the spin parameter $a$, reflecting an enhancement of the angular dependence of $\widetilde{T}(r,\, \theta)$. Nevertheless, even at $a = 0.998$, where the angular dependence is most pronounced among the cases examined, the maximum relative deviation from the reference value $\widetilde{T}(r,\, 0)$ remains below $4.5\%$. This result provides quantitative justification for neglecting the angular dependence of $\widetilde{T}(r,\, \theta)$ within the approximate analytical framework. Accordingly, the function $\widetilde{T}(r,\, \theta)$ can be approximated by a function of the radial coordinate $r$ alone over the investigated range of spin parameters.

According to Eq. (\ref{functionmrthetaa}), a function $\varepsilon (r\,, \theta)$ is defined as
\begin{equation}
  \varepsilon(r,\theta) \equiv \frac{a^2 \Delta_r \sin^2 \theta}{(r^2 + a^2)^2}\,.
\end{equation}
The properties of $\mathcal{M} (r,\, \theta;\, a)$ imply that $\varepsilon (r,\, \theta)$ attains the maximum value $\varepsilon_{\text{max}} = \varepsilon (r_{\text{max}},\, \pi / 2)$, which increases monotonically with the spin parameter $a$. For the largest spin parameter considered, $a = 0.998$, the maximum value is $\varepsilon_{\text{max}} \approx 0.0427$, indicating that $\varepsilon (r,\, \theta)$ is sufficiently small to serve as a perturbative expansion parameter over the investigated range of spin parameters. Accordingly, the function $T (r,\, \theta)$ can be expanded in powers of $\varepsilon (r,\, \theta)$ as
\begin{equation}\label{seriesexpansiont}
  T (r,\, \theta) \simeq T (r) + \mathcal{O} (\varepsilon) = \frac{4 (1 - \zeta) a^2 r^2}{\Delta_r (r^2 + a^2)^2}L^2 + \mathcal{O} (\varepsilon)\,.
\end{equation}
As shown in Eq. (\ref{seriesexpansiont}), the leading order term depends exclusively on the radial coordinate $r$, while the angular dependence appears only in the first and higher order corrections. The preceding analysis demonstrates that the radial dependence of $T(r,\, \theta)$ is dominant, providing a quantitative basis for retaining only the leading order term $T(r)$. This radial contribution can therefore be incorporated into the radial part of the Hamilton-Jacobi equation governing photon motion in the rotating Ho\v{r}ava spacetime, allowing an approximate separation of variables.

The Hamilton-Jacobi equation can therefore be approximately separated into two independent equations by introducing a separation constant $\mathcal{K}$ as
\begin{equation}\label{separatehjequation}
    \begin{split}
    & \Delta_r \left(\frac{dS_r}{dr} \right)^2 - \frac{\left[(r^2 + a^2) E - a L \right]^2}{\Delta_r} + \mu^2 r^2 + \frac{4 a r (\sqrt{\zeta} - 1)}{\Delta_r} E L + T (r) = - \mathcal{K}\,, \\
    & \left(\frac{dS_\theta}{d\theta} \right)^2 +\frac{\left(a \sin^2 \theta E - L \right)^2}{\sin^2 \theta} + a^2 \mu^2 \cos^2 \theta = \mathcal{K}\,.
  \end{split}
\end{equation}
The Carter constant $Q$ can be defined in terms of the separation constant $\mathcal{K}$ and the conserved quantities $E$ and $L$ as
\begin{equation}
  Q = \mathcal{K} - (L - a E)^2\,.
\end{equation}
The impact parameters $\lambda$ and $\eta$ characterizing photon motion in this spacetime are defined in terms of the Carter constant $Q$, conserved energy $E$, and angular momentum $L$ as
\begin{equation}\label{defimpactparameters}
  \lambda \equiv \frac{L}{E}\,, \qquad \eta \equiv \frac{Q}{E^2}\,.
\end{equation}

In the BL coordinates of the rotating Ho\v{r}ava spacetime, the contravariant four momentum of a particle can be expressed as $p^a = (p^t,\, p^r,\, p^\theta,\, p^\phi) = (\dot{t},\, \dot{r},\, \dot{\theta},\, \dot{\phi})$, where the dot denotes differentiation with respect to the affine parameter $\sigma$. Within the approximate separation framework, particle motion is characterized by the mass $\mu$, conserved energy $E$, angular momentum $L$, and Carter constant $Q$. The mass parameter $\mu$ is set to zero in the case of photon motion. By differentiating the particle action in Eq. (\ref{actionofparticle}) with respect to these constants and subsequently with respect to the affine parameter $\sigma$, the equations of motion along the four BL coordinates $(t,\, r,\, \theta,\, \phi)$ can be obtained as
\begin{equation}\label{equationofmotionbl}
  \begin{split}
    \frac{\rho^2}{E} p^t & = \frac{(r^2 + a^2)^2-2 a r \sqrt{\zeta} \lambda}{\Delta_r} - a^2 \sin^2 \theta\,,\\
    \frac{\rho^2}{E} p^r & = \sigma_r \sqrt{\mathcal{R} (r)}\,,\\
    \frac{\rho^2}{E} p^\theta & = \sigma_\theta \sqrt{\Theta(\theta)}\,,\\
    \frac{\rho^2}{E} p^\phi & = \frac{2 a r \sqrt{\zeta} - a^2 \lambda}{\Delta_r} + \lambda \csc^2 \theta - \frac{4 (\zeta - 1) a^2 r^2}{\Delta_r \Sigma^2} \lambda\,,
  \end{split}
\end{equation}
where the sign factors $\sigma_r, \sigma_\theta = \pm 1$ specify the radial and polar directions of the photon momentum components $p^r$ and $p^\theta$, respectively. The radial and angular potentials $\mathcal{R}(r)$ and $\Theta(\theta)$ governing photon motion in the rotating Ho\v{r}ava spacetime are defined as
\begin{equation}\label{radialangularpotential}
  \begin{split}
    \mathcal{R} (r) & = \left(r^2 + a^2 - a \lambda \right)^2 - \Delta_r \left[\eta + (\lambda - a)^2 \right] + 4 a r (1 - \sqrt{\zeta}) \lambda + \frac{4(\zeta - 1) a^2 r^2}{(r^2 + a^2)^2} \lambda^2\,,\\
    \Theta(\theta) & = \eta + a^2 \cos^2 \theta - \lambda^2 \cot^2 \theta\,.
  \end{split}
\end{equation}

The boundary of the black hole shadow is determined by photon trajectories that asymptotically approach unstable spherical photon orbits in the vicinity of the black hole. The construction of the shadow image therefore requires an analysis of the radial dynamics of photons associated with these orbits. The corresponding orbital radii can be determined by imposing the two conditions on the radial potential as
\begin{equation}\label{criticalcurvecondition}
  \mathcal{R} (\tilde{r}) = 0\,, \qquad \partial_r \mathcal{R} (\tilde{r}) = 0\,,
\end{equation}
where $\tilde{r}$ denotes the radius of an unstable spherical photon orbit in the vicinity of the black hole. The outer and inner boundaries of the radial range of unstable spherical photon orbits around the rotating Ho\v{r}ava black hole are denoted by $\tilde{r}_{+}$ and $\tilde{r}_{-}$, respectively. The explicit expressions for the impact parameters $\lambda$ and $\eta$ associated with these orbits can be derived by substituting Eq. (\ref{radialangularpotential}) into Eq. (\ref{criticalcurvecondition}) as
\begin{equation}\label{impactparameters}
  \begin{split}
    \lambda_s (\tilde{r}) = & \frac{\sqrt{\zeta} (\tilde{r}^2 - a^2) + s \dfrac{\Delta_{\tilde{r}}}{\tilde{r}^2 + a^2} \sqrt{\mathcal S(\tilde{r})}}{a\,\mathcal P(\tilde{r})}\,,\\
    \eta_s (\tilde{r}) = & \frac{\left[\tilde{r}^2 + a^2 - a \lambda_s (\tilde{r}) \right]^2 + 4 a \tilde{r} (1 - \sqrt{\zeta}) \lambda_s (\tilde{r}) + \dfrac{4 (\zeta - 1) a^2 \tilde{r}^2}{(\tilde{r}^2 + a^2)^2} \lambda_s^2 (\tilde{r})}{\Delta_{\tilde{r}}} - \left[\lambda_s (\tilde{r}) - a \right]^2\,,
  \end{split}
\end{equation}
where $\Delta_{\tilde{r}} \equiv \Delta_r|_{r = \tilde{r}}$ and $s = \pm 1$. The functions $\mathcal P (\tilde{r})$ and $\mathcal S (\tilde{r})$ appearing in Eq. (\ref{impactparameters}) are explicitly given by
\begin{equation}
  \begin{split}
    \mathcal{P} (\tilde{r}) & \equiv \tilde{r} - 1 + \frac{4 (\zeta - 1) \tilde{r} \left[- a^4 + a^2 \tilde{r} (\tilde{r} + 1) + \tilde{r}^3 (2 \tilde{r} - 3) \right]}{(\tilde{r}^2 + a^2)^3}\,, \\
    \mathcal{S} (\tilde{r}) & \equiv \tilde{r}^2 (\tilde{r}^2 + a^2)^2 + (\zeta - 1)(a^2 - 3 \tilde{r}^2)^2\,.
  \end{split}
\end{equation}
The impact parameters $\lambda_s (\tilde{r})$ and $\eta_s (\tilde{r})$ admit explicit expressions in terms of the critical photon orbit radius $\tilde{r}$. However, due to the complexity of $\eta_s (\tilde{r})$, a more convenient representation can be obtained by expressing $\eta_s (\tilde{r})$ in terms of $\lambda_s (\tilde{r})$.

The parameter $s = \pm 1$ in Eq. (\ref{impactparameters}) distinguishes the two algebraic branches of the photon impact parameters. According to Eq. (\ref{equationofmotionbl}), physically admissible photon trajectories are required to satisfy the condition $\Theta(\theta_o) \ge 0$ imposed by the angular potential. Numerical analysis indicates that this condition is satisfied only by the branch with $s = -1$. Accordingly, the impact parameters associated with unstable spherical photon orbits around the rotating Ho\v{r}ava black hole are given by $\lambda_{-}(\tilde{r})$ and $\eta_{-}(\tilde{r})$. For simplicity, these functions are denoted by $\lambda(\tilde{r})$ and $\eta(\tilde{r})$, respectively. In the limit $\zeta \rightarrow 1$, the photon impact parameters reduce to the Kerr expressions $\lambda_{\text{K}}(\tilde{r})$ and $\eta_{\text{K}}(\tilde{r})$ given by
\begin{equation}
  \begin{split}
    \lambda_{\text{K}} (\tilde{r}) = \frac{\tilde{r}^2 (3 - \tilde{r}) + a^2 (\tilde{r} + 1)}{a (\tilde{r} - 1)}\,, \qquad \eta_{\text{K}} (\tilde{r}) = \frac{\tilde{r}^3 \left[4 a^2 - \tilde{r} (\tilde{r} - 3 )^2 \right]}{a^2 (\tilde{r} - 1)^2}\,.
  \end{split}
\end{equation}
Furthermore, when $a = 0$, the rotating Ho\v{r}ava black hole reduces to the Schwarzschild black hole, and all unstable spherical photon orbits outside the event horizon lie on the photon sphere at $\tilde{r} = 3$. In this case, the photon impact parameters are not uniquely determined by the orbital radius $\tilde{r}$ and are instead constrained by the relation $\eta + \lambda^2 = 27$. In contrast, when $a \neq 0$, the explicit expressions for the photon impact parameters in Eq. (\ref{impactparameters}) allow the corresponding black hole shadow to be constructed within the approximate analytical framework.

\section{Construction of rotating Ho\v{r}ava black hole shadows using approximate analytical and numerical methods}\label{section3}

The shadow of a rotating Ho\v{r}ava black hole can be constructed analytically through approximate separation of the Hamilton-Jacobi equation governing photon motion. Alternatively, the numerical backward ray-tracing method provides an independent approach to constructing the shadow and is expected to yield more accurate results than the approximate analytical method. The accuracy of the analytical approximation can therefore be evaluated by comparing the corresponding shadow images. The construction of the black hole shadow also requires specifying the location of the observer, with the observer positioned either at spatial infinity or at a finite radial distance from the black hole. Accordingly, shadow images are constructed for both observer locations using the approximate analytical method and subsequently compared with the corresponding numerical results to assess the accuracy of the analytical construction.

\subsection{Construction on shadow images of rotating Ho\v{r}ava black holes at infinity}

The Bardeen coordinates $(x_{\text{B}},\, y_{\text{B}})$ are commonly employed to characterize the black hole shadow image on the image plane of an observer at spatial infinity. For an observer with an inclination angle $\theta_o$, these coordinates are given by
\begin{equation}\label{bardeencoordinate}
  x_{\text{B}} = - \frac{\lambda (\tilde{r})}{\sin \theta_o}\,, \qquad y_{\text{B}} = \sigma_{\theta} \sqrt{\Theta (\theta_{o})}\,.
\end{equation}
By substituting Eqs. (\ref{impactparameters}) and (\ref{radialangularpotential}) into Eq. (\ref{bardeencoordinate}), the shadow image of the rotating Ho\v{r}ava black hole can be constructed using the approximate analytical method. For comparison, the corresponding shadow image is independently constructed using the numerical backward ray-tracing method described in Appendix \ref{appendixa}.

Fig. \ref{fig2} presents the shadow images of rotating Ho\v{r}ava black holes constructed using the approximate analytical method and numerical backward ray-tracing on the Bardeen image plane of an observer at spatial infinity.
\begin{figure}[htbp]
    \centering
    \begin{subfigure}[b]{0.3\textwidth}
        \includegraphics[width=\textwidth]{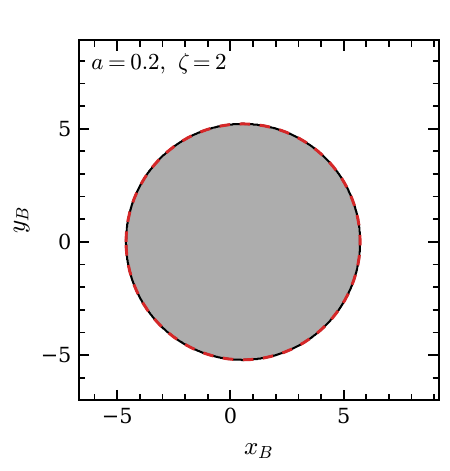}
        \caption{}
    \end{subfigure}
    \begin{subfigure}[b]{0.3\textwidth}
        \includegraphics[width=\textwidth]{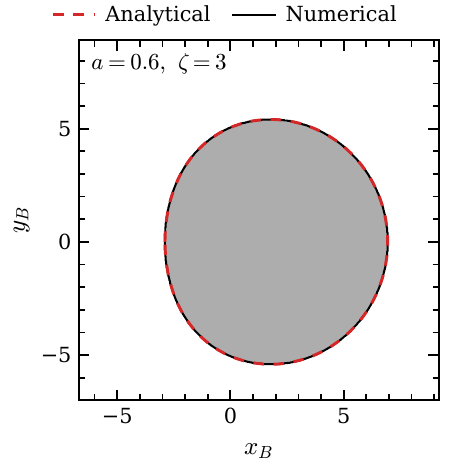}
        \caption{}
    \end{subfigure}
    \begin{subfigure}[b]{0.3\textwidth}
        \includegraphics[width=\textwidth]{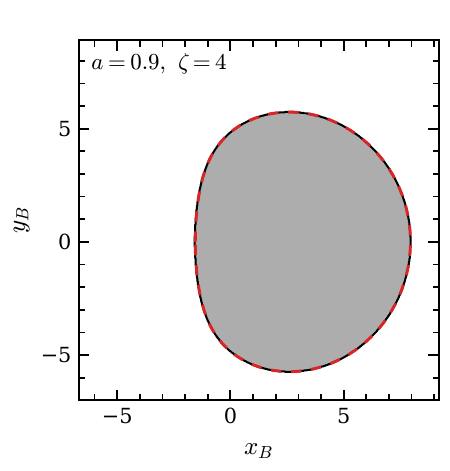}
        \caption{}
    \end{subfigure}
    \caption{Comparison of the shadow images of rotating Ho\v{r}ava black holes constructed at spatial infinity with a fixed observer inclination angle $\theta_o = 80^\circ$ for different combinations of the spin parameter $a = 0.2$, $0.6$, and $0.9$ and the LV factor $\zeta = 2$, $3$, and $4$. The red dashed and black solid curves represent the results obtained using the approximate analytical method and the numerical backward ray tracing method, respectively.}
    \label{fig2}
\end{figure}
The shadow images are presented for spin parameters $a = 0.2$, $0.6$, and $0.9$ and LV factors $\zeta = 2$, $3$, and $4$, with the observer inclination angle fixed at $\theta_o = 80^\circ$. The black solid curves represent the critical curves obtained from numerical backward ray-tracing, whereas the red dashed curves correspond to those constructed using the approximate analytical method. As illustrated in Fig. \ref{fig2}, the approximate analytical results are in excellent agreement with the numerical results for all considered values of $a$ and $\zeta$. With the numerical results serving as the reference, this agreement demonstrates the high accuracy of the approximate analytical method in reproducing the rotating Ho\v{r}ava black hole shadow images. Consequently, the approximate analytical method provides a reliable alternative to numerical calculations for investigating the properties of these shadow images. Furthermore, the agreement between the two methods supports the validity of the approximate separation of variables in the Hamilton-Jacobi equation governing photon motion in the rotating Ho\v{r}ava spacetime.

\subsection{Construction on shadow images of rotating Ho\v{r}ava black holes at finite distance}

The construction of the rotating Ho\v{r}ava black hole shadow image for an observer at a finite radial distance requires specifying an appropriate local reference frame. The zero angular momentum observer (ZAMO) frame is commonly adopted for this purpose. To establish the relationship between the shadow images observed at finite radial distances and those at spatial infinity, the correspondence between the imaging methods employed in the ZAMO frame and on the Bardeen image plane is first investigated.

The relationship between the photon projection coordinates on the ZAMO image plane and the corresponding Bardeen coordinates at spatial infinity is derived in Appendix \ref{zamobardeenrelation} and given by Eq. (\ref{coordinatesrelationbardeenzamo}). At leading order in the asymptotic expansion, Eq. (\ref{coordinatesrelationbardeenzamo}) can be expressed as
\begin{equation}\label{finalrelationbardeenzamo}
  \begin{split}
    \lim_{r \to \infty} \frac{2r}{\mathcal{L}} \left(x_{\rm Z},\, y_{\rm Z}\right) = \left(x_B,\, y_B \right)\,.
  \end{split}
\end{equation}
As demonstrated by Eq. (\ref{finalrelationbardeenzamo}), as the observer is displaced from a finite radial distance toward spatial infinity, the rescaled photon projection coordinates on the ZAMO image plane asymptotically approach the corresponding Bardeen coordinates. However, the original projection coordinates in the two coordinate systems are related through the scaling factor $2 r / \mathcal{L}$ rather than coinciding directly in the limit $r \rightarrow \infty$. Therefore, the physical significance of this scaling factor in establishing the correspondence between the two coordinate systems requires further clarification.

As demonstrated by the derivation of Eq. (\ref{coordinatesrelationbardeenzamo}) in Appendix \ref{zamobardeenrelation}, the numerical factor of $2$ in the scaling relation originates from the stereographic projection employed in defining the ZAMO image coordinates in the asymptotic limit $r \rightarrow \infty$. As the observer is displaced from a finite radial distance toward spatial infinity while the coordinate origin of the image plane remains fixed, the photon projection coordinates on the ZAMO image plane asymptotically approach the coordinate origin, indicating a contraction of the shadow image. The divergence of the radial coordinate $r$ in the scaling factor compensates for this contraction, ensuring that the rescaled projection coordinates remain finite in the limit $r \rightarrow \infty$. Accordingly, the black hole shadow image retains a finite extent in the rescaled coordinates at spatial infinity. Furthermore, for a projection scheme with the camera parameter $\mathcal{L} = 2$, the rescaled shadow image constructed in the ZAMO frame asymptotically approaches the corresponding shadow image expressed in Bardeen coordinates. Since the ZAMO image plane employs stereographic projection whereas the Bardeen image plane employs pinhole projection, the asymptotic relation between the corresponding photon projection coordinates establishes a mapping between these two projection schemes. This correspondence is valid only in the asymptotic limit of small angles and does not generally extend to finite observer distances, where the two projection schemes produce different image plane coordinates.

\begin{figure}[htbp]
    \centering
    \begin{subfigure}[b]{0.3\textwidth}
        \includegraphics[width=\textwidth]{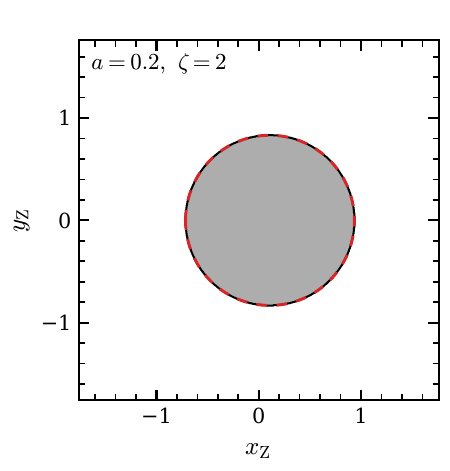}
        \caption{}
    \end{subfigure}
    \begin{subfigure}[b]{0.3\textwidth}
        \includegraphics[width=\textwidth]{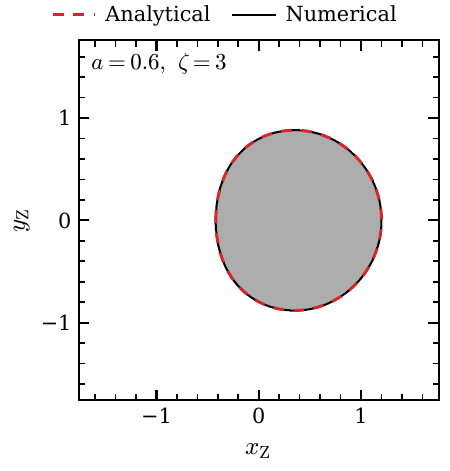}
        \caption{}
    \end{subfigure}
    \begin{subfigure}[b]{0.3\textwidth}
        \includegraphics[width=\textwidth]{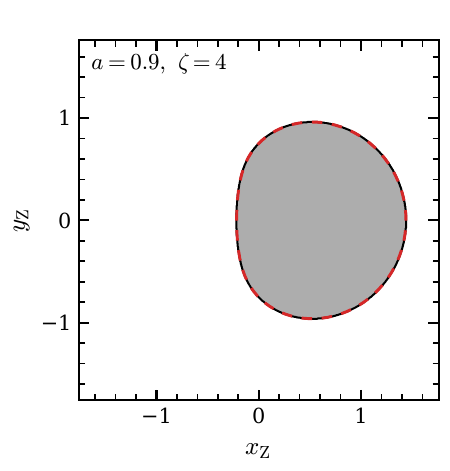}
        \caption{}
    \end{subfigure}
    \caption{Comparison of the shadow images of rotating Ho\v{r}ava black holes constructed at finite distance for different combinations of the spin parameter $a = 0.2$, $0.6$, and $0.9$ and the LV factor $\zeta = 2$, $3$, and $4$. The red dashed and black solid curves represent the results obtained using the approximate analytical method and the numerical backward ray-tracing method, respectively.}
    \label{fig3}
\end{figure}

The approximate analytical method and the numerical backward ray-tracing method detailed in Appendix \ref{zamoanalyticalnumericalmethod} are employed to construct the shadow images of rotating Ho\v{r}ava black holes on the ZAMO image plane at finite distances. The corresponding results are presented in Fig. \ref{fig3}.
For an observer located at $r_0 = 6$ with an inclination angle of $\theta_o = 80^\circ$, the shadow images are presented for spin parameters $a = 0.2$, $0.6$, and $0.9$ and LV factors $\zeta = 2$, $3$, and $4$. The black solid curves represent the critical curves obtained from numerical backward ray-tracing, whereas the red dashed curves correspond to those constructed using the approximate analytical method. As illustrated in Fig. \ref{fig3}, the approximate analytical results are in excellent agreement with the numerical results for all considered values of $a$ and $\zeta$. This agreement demonstrates the high accuracy of the approximate analytical method in reproducing the rotating Ho\v{r}ava black hole shadow images in the ZAMO frame at finite distances. Consequently, the approximate analytical method can also serve as a reliable alternative to numerical backward ray-tracing for constructing black hole shadow images at finite observer distances.

\section{Investigation of the shadow area of the rotating Ho\v{r}ava black hole}\label{section4}

The shadow images of rotating Ho\v{r}ava black holes have been constructed for observers at both finite distances and spatial infinity using the approximate analytical method and numerical backward ray-tracing. The excellent agreement between the approximate analytical and numerical results demonstrates that the approximate analytical method provides a reliable alternative to numerical calculations for investigating the properties of these shadow images. The black hole shadow is defined as the region enclosed by the critical curve on the image plane, and the shadow area is closely related to the parameters characterizing the black hole. Accordingly, it is necessary to calculate the shadow area and systematically investigate its dependence on the spin parameter $a$, the LV factor $\zeta$, and the observer inclination angle $\theta_o$.

\subsection{Shadow area of the rotating Ho\v{r}ava black hole shadow at infinity}

The investigation initially focuses on the shadow area of rotating Ho\v{r}ava black holes as determined from shadow images constructed at spatial infinity. The analysis proceeds with the calculation of the shadow area. The effects of variations in black hole parameters and observer configurations on the shadow area are subsequently examined through a systematic parameter study. This analysis provides a quantitative characterization of the shadow area at spatial infinity.

\subsubsection{Calculation of the shadow area at infinity}\label{}

A quantitative investigation of the shadow area of rotating Ho\v{r}ava black holes requires an appropriate method of calculation. Previous studies have commonly employed the average radius to characterize the overall size of black hole shadows \cite{Zhong:2021mty}. This approach requires determining the center of the black hole shadow and introducing a polar coordinate system $(\rho,\, \theta)$ with the origin located at the shadow center, where $\rho$ and $\theta$ denote the radial and angular coordinates, respectively. The average radius is defined by the angular average of $\rho(\theta)$ as
\begin{equation}
  \begin{split}
    \bar{r} = \frac{1}{2 \pi} \int_{0}^{2 \pi} \rho(\theta) d \theta\,.
  \end{split}
\end{equation}
However, irregular shadow geometries significantly complicate the analytical determination of the shadow center and the radial profile $\rho(\theta)$. These difficulties necessitate a more general approach to the calculation of the shadow area.

Since the black hole shadow corresponds to the region enclosed by the critical curve on the image plane, Green's theorem can be employed to calculate the shadow area. For an arbitrary simple closed curve $\mathcal{C}$ enclosing a planar region and parameterized by $t \in [t_0,\, t_1]$, the enclosed area can be expressed using Green's theorem as
\begin{equation}\label{definegreentheorem}
  \mathcal A = \oint_{\mathcal C} y \, d x = \left|\int_{t_0}^{t_1} y(t)\frac{dx(t)}{d t} \, d t \right|\,,
\end{equation}
where $x(t)$ and $y(t)$ denote the Cartesian coordinates of points on the closed curve $\mathcal{C}$ as functions of the parameter $t$. In the evaluation of the rotating Ho\v{r}ava black hole shadow area, the integration contour $\mathcal{C}$ is identified with the critical curve on the image plane, and the Bardeen coordinates $(x_{\text{B}},\, y_{\text{B}})$ are adopted as the Cartesian coordinates $(x,\, y)$. The parameterization of the critical curve in terms of the photon impact parameter $\lambda$ and the reflection symmetry of the shadow image with respect to the axis $y_{\text{B}} = 0$ lead to an integral expression for the shadow area given by
\begin{equation}\label{greentheoremlambda}
  \begin{split}
    \mathcal{A}_{\text{B}} = \frac{2}{\sin \theta_o} \int_{\lambda_-}^{\lambda_+} \sqrt{\Theta (\theta_o)} d \lambda\,,
  \end{split}
\end{equation}
where the integration limits $\lambda_-$ and $\lambda_+$ are determined by the two roots of the angular potential equation $\Theta(\theta_o) = 0$. The physically admissible values of the photon impact parameter are therefore restricted to $\lambda \in [\lambda_-,\, \lambda_+]$ by the requirement that $\Theta(\theta_o) \geq 0$. In the limit $a \rightarrow 0$, the rotating Ho\v{r}ava black hole reduces to the Schwarzschild black hole, and the shadow approaches the Schwarzschild configuration. The Bardeen coordinates and photon impact parameters associated with the critical curve satisfy $x_{\text{B}}^2 + y_{\text{B}}^2 = \eta + \lambda^2 = 27$, indicating that the shadow is circular with radius $r_{\text{Sch}} = 3 \sqrt{3}$ and area $\mathcal{A}_{\text{B},\, \text{Sch}} = 27 \pi$. The Schwarzschild shadow area is therefore adopted as a reference for quantifying deviations in the shadow area of rotating Ho\v{r}ava black holes, and the normalized shadow area is defined by the ratio of the rotating Ho\v{r}ava black hole shadow area to the Schwarzschild shadow area as
\begin{equation}\label{analyticalshadowarea}
  \begin{split}
    \mathcal{A}_{\text{NB},\, \text{analytical}} = \frac{\mathcal{A}_{\text{B}}}{\mathcal{A}_{\text{B},\, \text{Sch}}}\,.
  \end{split}
\end{equation}

\begin{figure}[htbp]
    \centering
    \includegraphics[width=0.5\textwidth]{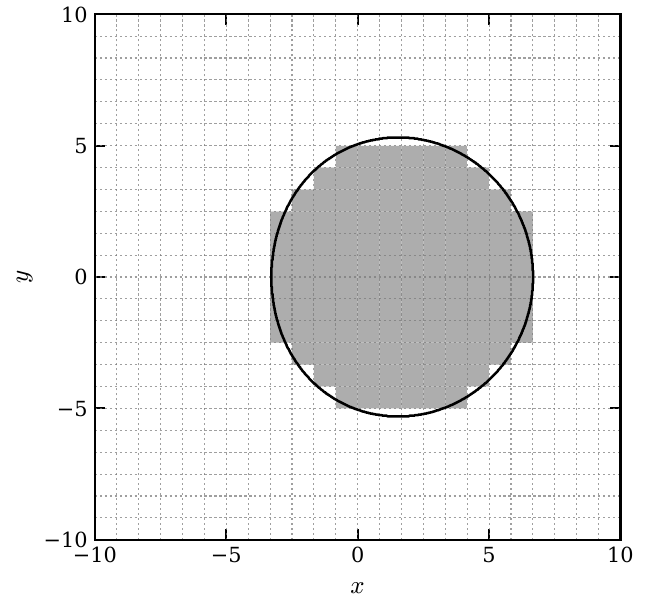}
    \caption{Schematic illustration of the numerical evaluation of the black hole shadow area. The gray dashed lines delineate the discretized grid on the image plane, while the black solid curve represents the critical curve obtained using the numerical backward ray tracing method. The cells shaded in gray represent the grid cells enclosed by the critical curve.}
    \label{fig4}
\end{figure}

In addition to the approximate analytical calculation, the shadow area of the rotating Ho\v{r}ava black hole is independently evaluated using the numerical backward ray-tracing method. The image plane is discretized into a grid, and the shadow area is determined by counting the pixels enclosed by the critical curve. The numerical procedure for calculating the shadow area is illustrated schematically in Fig. \ref{fig4}.
The red dashed curve in Fig. \ref{fig4} represents the critical curve of the rotating Ho\v{r}ava black hole obtained using numerical backward ray-tracing. The gray shaded grid cells indicate the region enclosed by the critical curve on the image plane. For a square image plane with side length $L_{\text{scr}}$ discretized into $n$ grid cells along each axis, the total area of the image plane is $\mathcal{A}_{\text{scr}} = L_{\text{scr}}^2$, and the total number of grid cells is $N_{\text{scr}} = n^2$. The number of grid cells enclosed by the critical curve is denoted by $N_{\text{sh}}$. The numerical shadow area of the rotating Ho\v{r}ava black hole normalized by the Schwarzschild shadow area can be expressed as
\begin{equation}\label{numericalshadowarea}
  \mathcal{A}_{\text{NB},\, \text{numerical}} = \frac{N_{\text{sh}}}{N_{\rm scr}} \frac{\mathcal{A}_{\text{scr}}}{\mathcal{A}_{\text{B},\, \text{Sch}}}\,.
\end{equation}

\begin{figure}[t]
    \centering
    \begin{subfigure}[b]{0.48\textwidth}
        \centering
        \includegraphics[width=\textwidth]{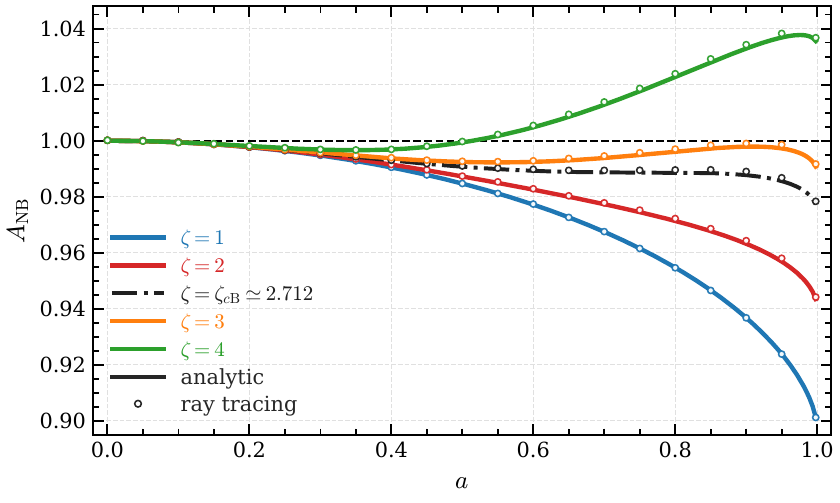}
        \caption{}
        \label{}
    \end{subfigure}
    \hfill
    \begin{subfigure}[b]{0.48\textwidth}
        \centering
        \includegraphics[width=\textwidth]{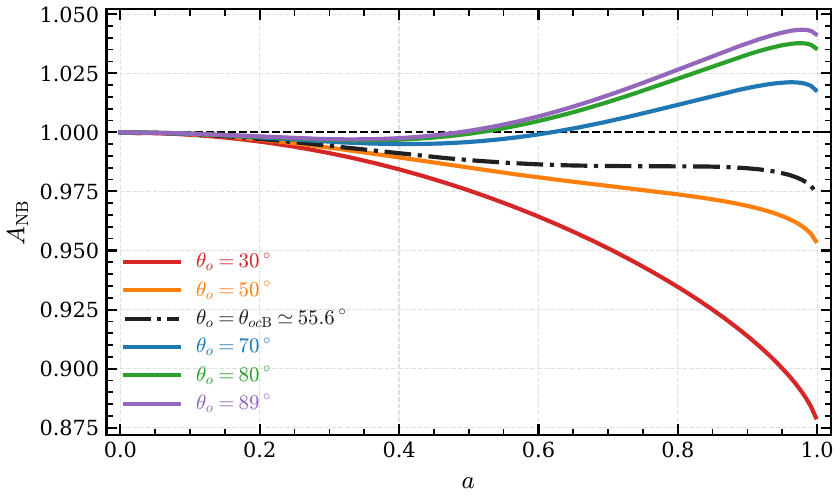}
        \caption{}
        \label{}
    \end{subfigure}
    \caption{The normalized shadow area of the rotating Ho\v{r}ava black hole at spatial infinity is plotted as a function of the spin parameter $a$ for different values of the LV factor $\zeta$ and the observer inclination angle $\theta_o$. The solid curves and open circles represent the approximate analytical and numerical results, respectively. Panel (a) illustrates the dependence on $a$ for different values of $\zeta$ at a fixed observer inclination angle $\theta_o = 80^\circ$, whereas panel (b) illustrates the corresponding dependence for different observer inclination angles $\theta_o$ at a fixed value of the LV factor $\zeta = 4$.}
    \label{fig5}
\end{figure}

The normalized shadow area of the rotating Ho\v{r}ava black hole can be calculated using the approximate analytical and numerical methods according to Eqs. (\ref{analyticalshadowarea}) and (\ref{numericalshadowarea}), respectively. The dependence of the normalized shadow area on the spin parameter $a$ for different values of the LV factor $\zeta$ and the observer inclination angle $\theta_0$ is illustrated in Fig. \ref{fig5}.
The dependence of the normalized shadow area of rotating Ho\v{r}ava black holes on the spin parameter $a$ for an observer inclination angle $\theta_o = 80^\circ$ and LV factors $\zeta = 1$, $2$, $3$, and $4$ is illustrated in Fig. \ref{fig5} (a). The normalized shadow areas calculated using the approximate analytical method are plotted as solid curves, whereas the numerical backward ray-tracing results are represented by open circles. The approximate analytical and numerical results exhibit excellent agreement for all considered values of $a$ and $\zeta$, demonstrating the high accuracy of the approximate analytical method and confirming that this method provides a reliable alternative to numerical calculations of the rotating Ho\v{r}ava black hole shadow area.

The horizontal dashed line at $\mathcal{A}_{\text{NB}} = 1$ in Fig. \ref{fig5} represents the normalized Schwarzschild black hole shadow area. In the nonrotating limit $a = 0$, the rotating Ho\v{r}ava black hole reduces to the Schwarzschild black hole, and the normalized shadow area equals unity independently of the LV factor $\zeta$. For $\zeta = 1$, the rotating Ho\v{r}ava black hole reduces to the Kerr black hole, and the blue solid curve characterizes the dependence of the Kerr shadow area on the spin parameter $a$. The Kerr shadow area exhibits a monotonic decrease from the Schwarzschild value as the spin parameter $a$ increases. For $\zeta = 2$, the shadow area similarly decreases monotonically with increasing $a$ and remains below the Schwarzschild value for nonzero spin, although the reduction is less pronounced than in the Kerr case. In contrast, when the LV factor exceeds the critical value $\zeta_{\text{cB}} \simeq 2.712$, the shadow area initially decreases to a minimum before exhibiting an anomalous increase with increasing $a$. This distinctive behavior differs from that of the Kerr black hole and rotating Ho\v{r}ava black holes with $\zeta < \zeta_{\text{cB}}$. For $\zeta > \zeta_{\text{cB}}$, the shadow area reaches a maximum at $a > 0.9$ and subsequently decreases slightly as the spin parameter approaches $a = 1$. The increase in the shadow area with increasing spin becomes progressively more pronounced as the LV factor $\zeta$ increases, with the strongest enhancement occurring at $\zeta = 4$ among the values considered.

The dependence of the rotating Ho\v{r}ava black hole shadow area on the spin parameter $a$ for different observer inclination angles $\theta_o$ at a fixed LV factor $\zeta = 4$ is illustrated in Fig. \ref{fig5} (b). The results reveal a critical inclination angle $\theta_{\text{ocB}} = 55.6^\circ$ that separates two distinct regimes in the evolution of the shadow area with increasing $a$. For $\theta_o < \theta_{\text{ocB}}$, the shadow area decreases monotonically with increasing $a$, although the reduction becomes progressively less pronounced as $\theta_o$ increases. For $\theta_o > \theta_{\text{ocB}}$, the monotonic decrease observed at lower inclinations is replaced by a distinct evolution in which the shadow area decreases to a minimum before increasing to a maximum and declining slightly as $a$ approaches unity. Moreover, the anomalous increase in the shadow area becomes progressively more substantial with increasing $\theta_o$. These results demonstrate that the shadow area of rotating Ho\v{r}ava black holes can exhibit an anomalous increase with increasing spin when $\theta_o > \theta_{\text{ocB}}$. The results presented in Figs. \ref{fig5} (a) and (b) demonstrate that an anomalous increase in the shadow area with increasing $a$ emerges when the LV factor $\zeta$ and the observer inclination angle $\theta_o$ exceed the corresponding critical values.

\subsubsection{Discussions of the anomalous increase of the shadow area at infinity}\label{}

The excellent agreement between the approximate analytical and numerical results provides a solid foundation for analytically investigating the origin of the anomalous increase in the shadow area of rotating Ho\v{r}ava black holes with increasing spin parameter $a$. However, the shadow area integral derived from Green's theorem cannot be evaluated analytically to obtain a closed-form expression. This limitation can be addressed by introducing a series expansion of the shadow area, which permits a systematic perturbative evaluation of the underlying mechanism responsible for the anomalous increase. Since this anomalous behavior occurs as the spin parameter $a$ increases, it is natural to adopt $a$ as the expansion parameter. Furthermore, the restricted range $a \in [0,\, 1]$ provides an additional motivation for this choice in view of the convergence of the resulting series expansion. In Eq. (\ref{greentheoremlambda}), the critical curve is parameterized by the photon impact parameter $\lambda$. The series expansion of the photon impact parameter $\eta$ can therefore be expressed as
\begin{equation}\label{tretaalambdaexpanda}
  \begin{split}
    \eta (a\,, \lambda) = 27 - \lambda^2 + a \eta_1 (\lambda) + a^2 \eta_2 (\lambda) + \mathcal O (a^3)\,.
  \end{split}
\end{equation}
The explicit expressions for the first- and second-order expansion coefficients can be obtained from Eq. (\ref{criticalcurvecondition}) as
\begin{equation}
  \begin{split}
    \eta_1 (\lambda) = - 4\sqrt{\zeta} \lambda\,, \qquad \eta_2 (\lambda) = \frac{5 - 8 \zeta}{27} \lambda^2 - 4\,.
  \end{split}
\end{equation}
Furthermore, substituting the expansion in Eq. (\ref{tretaalambdaexpanda}) and the angular potential $\Theta$ from Eq. (\ref{radialangularpotential}) into Eq. (\ref{greentheoremlambda}) leads to the series expansion of the rotating Ho\v{r}ava black hole shadow area as
\begin{equation}\label{shaodwareaexpansion}
  \begin{split}
    & \mathcal A_\infty (a\,, \zeta\,; \theta_o)\\
    = & 27\pi + \frac{\pi \sin^2 \theta_o}{4} \left[48 \zeta - 48 \zeta + \frac{4 (\cos^2 \theta_o - 4)}{\sin^2 \theta_o} + 10 \right] a^2\\
    & + \frac{\pi}{216} \left[144 \zeta^2 \sin^4 \theta_o + 4 \zeta (37 \cos^2 \theta_o - 49) \sin^2 \theta_o+ 19 \cos^4 \theta_o - 86 \cos^2 \theta_o - 29 \right] a^4\\
    & + \mathcal O (a^6)\\
    = & c_0 + c_2 \left(\theta_0 \right) a^2 + c_4 \left(\zeta\,, \theta_o \right) a^4 + \mathcal O (a^6)\,,
  \end{split}
\end{equation}
where the coefficients of the series expansion are given by
\begin{equation}\label{coeffinseriesexpan}
  \begin{split}
    & c_0 = 27 \pi\,, \qquad c_2 \left(\theta_o \right) = - \frac{3 \pi}{2} \left(1 + \cos^2 \theta_o \right)\,,\\
    & c_4 \left(\zeta\,, \theta_o \right) = \frac{2 \pi}{3} \zeta^2 \sin^4 \theta_o + \frac{\pi}{54} \zeta (37 \cos^2 \theta_o - 49) \sin^2 \theta_o + \frac{\pi}{216} \left(19 \cos^4 \theta_o - 86 \cos^2 \theta_o - 29 \right)\,.
  \end{split}
\end{equation}

As shown in Eq. (\ref{shaodwareaexpansion}), the series expansion of the black hole shadow area contains only even powers of the spin parameter $a$. This property arises from the invariance of the shadow area under the transformation $a \rightarrow -a$, which produces a mirror reflection of the critical curve without changing the enclosed area. The zeroth-order coefficient $c_0$ represents the Schwarzschild black hole shadow area, consistent with the static limit of the rotating Ho\v{r}ava black hole at $a = 0$. The second-order coefficient $c_2 \left(\theta_o \right)$ depends exclusively on the observer inclination angle $\theta_o$ and is independent of the LV factor $\zeta$. As established in the derivation of Eq. (\ref{shaodwareaexpansion}), the contributions to the second-order coefficient that depend on $\zeta$ cancel exactly. Consequently, $c_2 \left(\theta_o \right)$ coincides with the second-order coefficient for the Kerr black hole, demonstrating that the rotating Ho\v{r}ava and Kerr black hole shadow areas coincide through second order in $a$. This relation can be expressed as $c_2 \left(\theta_o \right) = c_2^{\text{Kerr}} \left(\theta_o \right)$. The fourth-order coefficient $c_4 (\zeta,\, \theta_o)$ in Eq. (\ref{coeffinseriesexpan}) can be decomposed as
\begin{equation}\label{coefficientc4}
  \begin{split}
    c_4(\zeta,\theta_o) = c_4^{\text{Kerr}} (\theta_o) + c_4^{\text{Ho\v{r}ava}} (\zeta\,, \theta_o)\,,
  \end{split}
\end{equation}
where
\begin{equation}
  \begin{split}
    & c_4^{\text{Kerr}} (\theta_o) = \frac{\pi}{72} \left(5 \cos^4 \theta_o - 10 \cos^2 \theta_o -27 \right)\,,\\
    & c_4^{\text{Ho\v{r}ava}} (\zeta,\, \theta_o) = \frac{\pi}{54} (\zeta - 1) \sin^2 \theta_o \left[36 (\zeta + 1) \sin^2 \theta_o + 37 \cos^2 \theta_o - 49 \right]\,.
  \end{split}
\end{equation}
The coefficient $c_4^{\text{Kerr}}(\theta_o)$ represents the fourth-order coefficient in the series expansion of the Kerr black hole shadow area, whereas $c_4^{\text{Ho\v{r}ava}}(\zeta,\,\theta_o)$ characterizes the correction to this coefficient arising from the LV factor in the rotating Ho\v{r}ava spacetime. The vanishing of this correction in the limit $\zeta \rightarrow 1$ recovers the Kerr value of the fourth-order coefficient. The dependence of the rotating Ho\v{r}ava black hole shadow area on the LV factor $\zeta$ does not arise until fourth order in $a$, as demonstrated by Eq. (\ref{coefficientc4}). Accordingly, an expansion through at least fourth order in $a$ is necessary to account for the dependence of the shadow area on the LV factor.

\begin{figure}[t]
  \centering
  \includegraphics[width=0.5\linewidth]{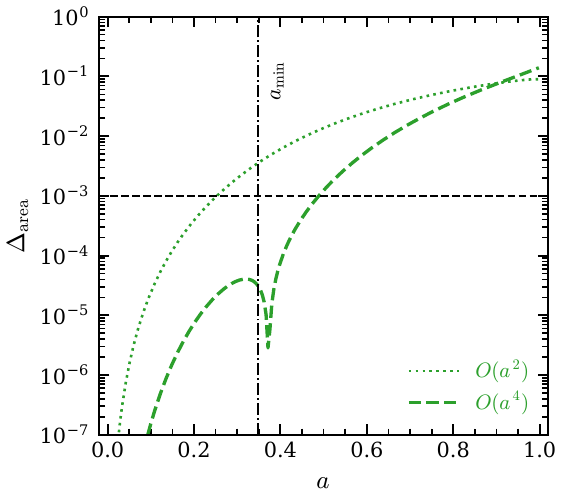}
  \caption{The relative deviation $\Delta\_{\text{area}}$ of the series approximations to the rotating Ho\v{r}ava black hole shadow area as a function of the spin parameter $a$ for $\zeta=4$ and $\theta_o=80^\circ$. The dotted and dashed curves represent the relative deviations associated with the second- and fourth-order truncations, respectively. The horizontal dashed line represents the reference value $\Delta_{\text{area}}=10^{-3}$, whereas the vertical dash dot line indicates the spin parameter $a_{\text{min}}$ corresponding to the minimum of the shadow area and the onset of the anomalous increase.
}
  \label{fig6}
\end{figure}

The selection of an appropriate truncation order for the series expansion is essential for investigating the anomalous increase in the rotating Ho\v{r}ava black hole shadow area with increasing spin parameter $a$. Since the LV factor contributes to the shadow area in the fourth-order term of the series expansion, the accuracy of the approximations can be assessed by comparing the shadow areas calculated from the expansions truncated at second- and fourth-order with the result obtained from the full expression for the shadow area. The discrepancies between the truncated expansions and the full result are quantified by the relative deviation defined as
\begin{equation}\label{relativedeviationsarea}
  \begin{split}
    \Delta_{\text{area}} = \frac{\left|\mathcal{A}_{\text{trunc}} - \mathcal{A}_{\text{B}} \right|}{\mathcal{A}_{\text{B}}}\,,
  \end{split}
\end{equation}
where $\mathcal{A}_{\text{trunc}}$ denotes the black hole shadow area obtained from the series expansion truncated at second or fourth order, whereas $\mathcal{A}_{\text{B}}$ represents the shadow area calculated from the full analytical expression. The relative deviations of the second- and fourth-order truncated shadow areas from the shadow area obtained using the full analytical expression are evaluated according to Eq. (\ref{relativedeviationsarea}). The dependence of these relative deviations on the spin parameter $a$ is illustrated in Fig. \ref{fig6} with the LV factor and the observer inclination angle fixed at $\zeta = 4$ and $\theta_o = 80^\circ$, respectively.
The dotted and dashed curves in Fig. \ref{fig6} represent the relative deviations of the second- and fourth-order approximations to the black hole shadow area from the area calculated using the full analytical expression, respectively. For $a \lesssim 0.3$, the relative deviations of both approximations remain below $10^{-3}$. As $a$ increases from approximately $0.3$ to $0.5$, the relative deviation of the second-order approximation rises above $10^{-3}$ and increases progressively with $a$, whereas that of the fourth-order approximation remains below $10^{-3}$. When $a$ exceeds approximately $0.5$, the relative deviation of the fourth-order approximation also exceeds $10^{-3}$ and increases with $a$. Furthermore, the shadow area curves exhibiting anomalous behavior in Fig. \ref{fig5} initially decrease to a local minimum before increasing with $a$. This minimum signifies the transition to an anomalous increase in the shadow area. The numerical analysis identifies the minimum at $a_{\text{min}} = 0.349$ for an LV factor $\zeta = 4$ and an observer inclination angle $\theta_o = 80^\circ$, as indicated by the vertical black dash dot line in Fig. \ref{fig6}. In the vicinity of $a_{\text{min}}$, the relative deviation of the fourth-order truncated expansion decreases sharply to approximately $10^{-6}$. The fourth-order expansion therefore accurately reproduces the rotating Ho\v{r}ava black hole shadow area in a neighborhood of $a_{\text{min}}$ and provides an analytical foundation for investigating the anomalous increase within this parameter range.

The validity of the fourth-order truncated expression for the rotating Ho\v{r}ava black hole shadow area can be further assessed analytically. The derivative of the shadow area expression in Eq. (\ref{shaodwareaexpansion}) with respect to $a$ is obtained as
\begin{equation}
  \frac{d \mathcal{A}_{\text{B}}}{d a} = 2 c_2 (\theta_o) a + 4 c_4 (\zeta\,, \theta_o) a^3\,.
\end{equation}
The spin parameter $a{\text{tmin}}$ for which the fourth-order truncated expression for the shadow area attains a local minimum is determined as
\begin{equation}\label{atmin}
  a_{\text{tmin}} = \sqrt{\frac{- c_2 (\theta_o)}{2 c_4 (\zeta\,, \theta_o)}} =\sqrt{\frac{3 \pi (1 + \cos^2 \theta_o)}{4 c_4 (\zeta\,, \theta_o)}}\,.
\end{equation}
The second-order coefficient in Eq. (\ref{coeffinseriesexpan}) satisfies $c_2 (\theta_o) < 0$ for all allowed observer inclination angles $\theta_o$. The fourth-order coefficient $c_4 (\zeta\,, \theta_o)$ is found numerically to be either negative or positive. A real positive solution for $a_{\text{tmin}}$ is obtained from Eq. (\ref{atmin}) only when $c_4 (\zeta\,, \theta_o) > 0$. The results for $c_4 (\zeta\,, \theta_o)$, $a_{\text{tmin}}$, and the variation of the black hole shadow area with increasing spin parameter $a$ are summarized in Table \ref{table2} for different LV factors at $\theta_o = 80^\circ$.
\begin{table}[t]
\centering
\caption{Summary of the fourth-order coefficient $c_4 (\zeta\,, \theta_o)$, the spin parameter $a_{\text{tmin}}$ corresponding to the minimum of the fourth-order approximation to the shadow area, and the qualitative behavior of the shadow area over the spin range $a \in [0,\, 1)$ for different values of $\zeta$ at a fixed observer inclination angle $\theta_o=80^\circ$.}
\begin{tabular}{cccc}
\hline
$\zeta$ & $c_4 (\zeta\,, \theta_o)$ & $a_{\text{tmin}}$ & behavior in $0 \le a < 1$ \\
\hline
$1$ & $-1.191$ & -- & monotonic decrease \\
$2$ & $+2.017$ & $1.097$ & monotonic decrease \\
$4$ & $+20.25$ & $0.346$ & minimum, then increase \\
\hline
\label{table2}
\end{tabular}
\end{table}
As shown in Table \ref{table2}, the fourth-order coefficient satisfies $c_4 (\zeta\,, \theta_o) < 0$ for the Kerr solution recovered at $\zeta = 1$, and the shadow area decreases monotonically with increasing spin parameter $a$ within the fourth-order approximation. The coefficient satisfies $c_4 (\zeta\,, \theta_o) > 0$ for $\zeta = 2$, whereas the minimum predicted by Eq. (\ref{atmin}) occurs at $a_{\text{tmin}} > 1$ outside the allowed interval $a \in [0,\, 1]$. The truncated shadow area therefore remains monotonically decreasing throughout this interval. By comparison, Eq. (\ref{atmin}) gives $a_{\text{tmin}} = 0.346$ at $\zeta = 4$, in excellent agreement with the numerical result $a_{\text{min}} = 0.349$. These results indicate that the fourth-order expansion accurately predicts the spin parameter at the shadow area minimum and provides an analytical framework for investigating the subsequent anomalous increase.

\begin{figure}[t]
  \centering
  \includegraphics[width=0.5\linewidth]{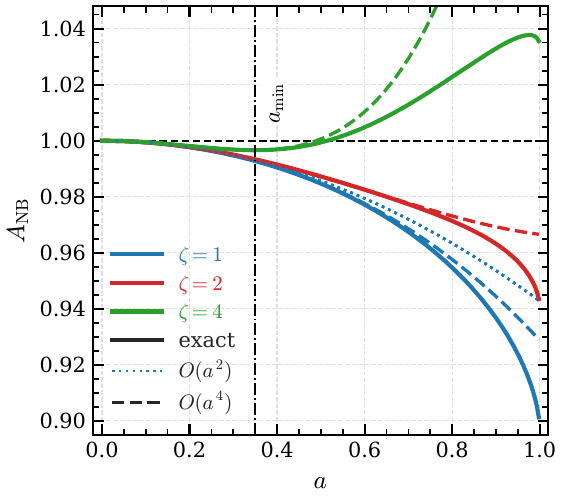}
  \caption{The rotating Ho\v{r}ava black hole shadow area at spatial infinity is calculated as a function of the spin parameter $a$ within the approximate analytical framework for $\zeta=1$, $2$, and $4$ at a fixed observer inclination angle $\theta_o=80^\circ$. The solid curves represent the full analytical expression for the black hole shadow area, whereas the dotted and dashed curves are obtained from the series expansions truncated at second- and fourth- order with respect to the spin parameter, respectively.}
  \label{fig7}
\end{figure}

The rotating Ho\v{r}ava black hole shadow area is calculated as a function of the spin parameter $a$ from the full analytical expression and the expansions truncated at second and fourth order. The resulting curves are compared in Fig. \ref{fig7} for an observer inclination angle $\theta_o = 80^\circ$ and LV factors $\zeta=1$, $2$, and $4$.
The solid, dotted, and dashed curves in Fig. \ref{fig7} illustrate the variation of the black hole shadow area with increasing spin parameter $a$, as calculated from the full analytical expression and the second- and fourth-order truncated expansions, respectively. In the static limit $a = 0$, the rotating Ho\v{r}ava black hole shadow areas in Fig. \ref{fig7} coincide with the Schwarzschild shadow area for all considered LV factors. This behavior arises from the fact that the leading-order term $c_0$ in Eq. (\ref{shaodwareaexpansion}) determines the shadow area in this limit and reproduces the Schwarzschild result independently of $\zeta$ and $\theta_o$, consistent with the Schwarzschild limit of the rotating Ho\v{r}ava black hole. As $a$ increases, higher-order terms in Eq. (\ref{shaodwareaexpansion}) contribute increasingly to the shadow area. For $\zeta = 1$, the expansion recovers the Kerr result, and the decrease in the shadow area over the approximate range $a \in [0,\, 0.5]$ is primarily described by the second-order term. In the regime $a > 0.5$, the discrepancy between the second-order approximation and the full analytical result increases, and the negative fourth-order term accounts for an additional reduction in the Kerr shadow area. In the case of $\zeta = 2$, the initial decrease is similarly described primarily by the second-order term over the approximate range $a \in [0,\, 0.4]$. Since $c_2 (\theta_o)$ coincides with the Kerr coefficient and is independent of $\zeta$, the second-order approximations for the different LV factors coincide on the same dotted curve. Within the parameter range $a > 0.4$, the positive fourth-order correction increasingly counteracts the negative second-order contribution, resulting in a less pronounced decrease in the shadow area than in the Kerr case. With the LV factor set to $\zeta = 4$, the second-order approximation describes the initial decrease over the approximate range $a \in [0,\, 0.2]$. The positive fourth-order correction increases in magnitude for $a > 0.2$ and raises the shadow area relative to the Kerr result. As $\zeta$ increases, the second-order approximation agrees closely with the full analytical result over an increasingly restricted range of spin parameters. The fourth-order correction also becomes significant at smaller values of $a$, enhancing the shadow area relative to the Kerr case.

\begin{figure}[t]
  \centering
  \includegraphics[width=0.5\linewidth]{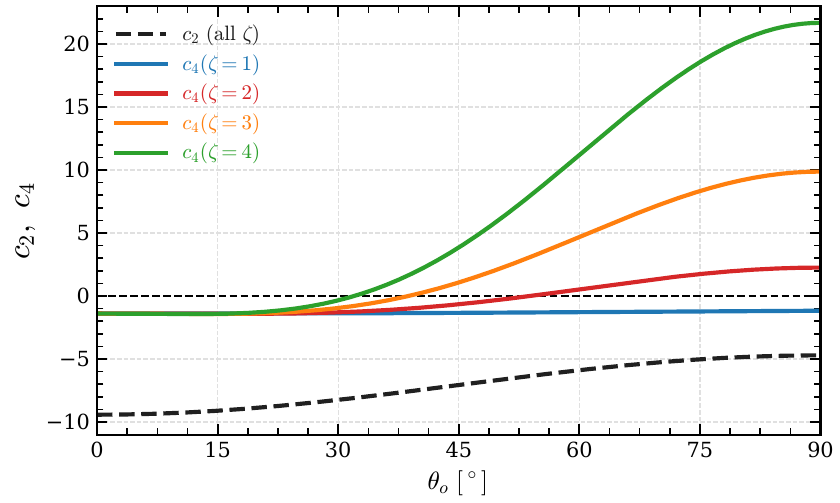}
  \caption{The second-order coefficient $c_2$ and the fourth-order coefficient $c_4$ in the black hole shadow area expansion are plotted as functions of the observer inclination angle $\theta_o$, with the latter evaluated for $\zeta=1$, $2$, $3$ and $4$.}
  \label{fig8}
\end{figure}

The dependence of the second-order coefficient $c_2 (\theta_o)$ on the observer inclination angle $\theta_o$ and that of the fourth-order coefficient $c_4 (\zeta,\, \theta_o)$ on both $\theta_o$ and the LV factor $\zeta$ can be further investigated on the basis of Eq. (\ref{coeffinseriesexpan}). These parameter dependencies are illustrated in Fig. \ref{fig8}.
As illustrated in Fig. \ref{fig8}, the second-order coefficient $c_2 (\theta_o)$ remains negative and approaches zero as the observer inclination angle $\theta_o$ increases, indicating that the second-order contribution consistently reduces the black hole shadow area. This behavior is consistent with the analytical expression for $c_2 (\theta_o)$ in Eq. (\ref{coeffinseriesexpan}). At inclination angles below $30^\circ$, the fourth-order coefficient $c_4 (\theta_o,\, \zeta)$ remains nearly identical and negative for different LV factors, indicating that the fourth-order contribution also reduces the shadow area independently of $\zeta$. At inclination angles above $30^\circ$, however, the LV factor $\zeta$ significantly alters the dependence of $c_4$ on $\theta_o$. At $\zeta = 1$, the fourth-order coefficient remains approximately constant and negative, contributing to a further reduction in the Kerr black hole shadow area. When $\zeta = 2$, the coefficient changes sign and attains positive values at inclination angles above $60^\circ$, implying that the fourth-order contribution changes from reducing to enhancing the shadow area. As $\zeta$ increases further, this transition occurs at progressively smaller inclination angles. An increase in $\zeta$ enhances the angular variation of $c_4$, with the largest positive change in the coefficient observed at $\zeta = 4$ among the values considered. These results indicate that the positive fourth-order contribution can overcome the decreasing trend induced by the negative second-order term at sufficiently large observer inclination angles and high LV factors, thereby accounting for the anomalous increase in the rotating Ho\v{r}ava black hole shadow area with increasing spin parameter $a$.

\subsection{Shadow area of the rotating Ho\v{r}ava black hole shadow at finite distances}
 
The anomalous increase in the rotating Ho\v{r}ava black hole shadow area with increasing spin parameter $a$ has been investigated analytically for observers at spatial infinity. The analytical results establish that the anomalous increase is primarily governed by the fourth-order term in the shadow area expansion, whereas all contributions involving the LV factor $\zeta$ cancel exactly at second order, leaving the second-order coefficient identical to the corresponding coefficient in the Kerr black hole shadow area expansion. Since the shadow at spatial infinity provides the asymptotic reference for shadows constructed at finite distances, it is necessary to further investigate the shadow area at finite distances and establish the relationship between the shadow areas at finite distances and spatial infinity.

\subsubsection{Calculation of the shadow area at finite distances}

The shadow area of the rotating Ho\v{r}ava black hole at a finite observer distance can be evaluated analogously through the area integral derived from Green's theorem in Eq. (\ref{definegreentheorem}). The Cartesian coordinates $(x,\, y)$ in this integral are identified with the photon projection coordinates $(x_\text{Z},\, y_\text{Z})$ on the ZAMO image plane, as defined in Eq. (\ref{specificexpzamocoordinates}). With the critical curve parameterized by the photon impact parameter $\lambda$, the shadow area on the ZAMO image plane can be expressed as
\begin{equation}\label{greentheoremzamo}
  \mathcal A_{\text{Z}} = 2 \left|\int_{\lambda_-}^{\lambda_+} y_{{\text{Z}}\,, +}(\lambda) \frac{d x_{\text{Z}} (\lambda)}{d \lambda} d \lambda \right|\,,
\end{equation}
The ZAMO projection at finite observer distances and the Bardeen projection at spatial infinity exhibit an asymptotic correspondence described by Eq. (\ref{finalrelationbardeenzamo}), which also specifies the relationship between the corresponding photon projection coordinates through the scaling factor $2r/\mathcal{L}$. To mitigate the influence of this scaling factor on the shadow area analysis and facilitate comparison with the Schwarzschild result, the rotating Ho\v{r}ava black hole shadow area at finite observer distances is normalized by the Schwarzschild shadow area following the procedure adopted at spatial infinity. The normalized shadow area is expressed formally as
\begin{equation}
  \begin{split}
    \mathcal{A}_{\text{NZ},\, \text{analytical}} = \frac{\mathcal{A}_\text{Z}}{\mathcal{A}_{\text{Z},\, \text{Sch}}}\,,
  \end{split}
\end{equation}
where $\mathcal{A}_{\text{Z},\, \text{Sch}}$ denotes the Schwarzschild black hole shadow area on the ZAMO image plane at a finite observer distance, obtained by evaluating the rotating Ho\v{r}ava black hole shadow area in the limit $a = 0$. The shadow area at finite observer distances can also be calculated numerically using the procedure schematically illustrated in Fig. \ref{fig4}. In a form analogous to Eq. (\ref{numericalshadowarea}), the numerical shadow area of the rotating Ho\v{r}ava black hole normalized by $\mathcal{A}_{\text{Z},\, \text{Sch}}$ can be expressed as
\begin{equation}
  \mathcal{A}_{\text{NZ},\, \text{numerical}} = \frac{N_{\text{sh}}}{N_{\rm scr}} \frac{\mathcal{A}_{\text{scr}}}{\mathcal{A}_{\text{Z}\,, \text{Sch}}}\,.
\end{equation}

\begin{figure}[htbp]
    \centering
    \begin{subfigure}[b]{0.48\textwidth}
        \includegraphics[width=\textwidth]{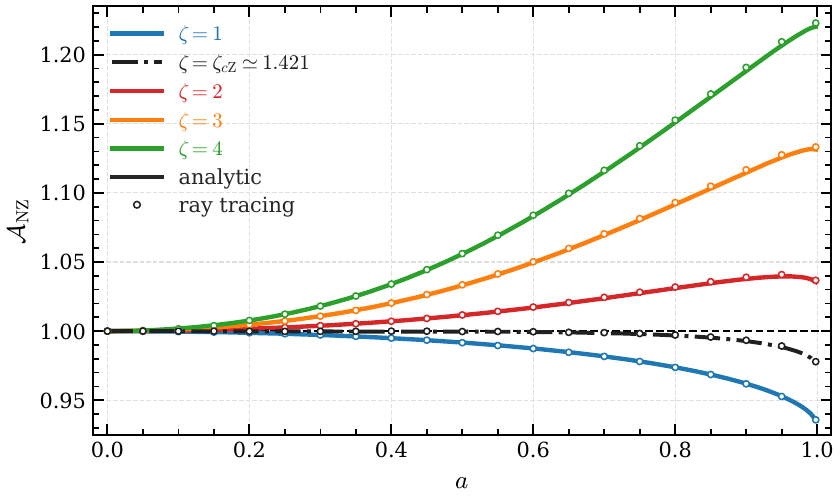}
        \caption{}
    \end{subfigure}
    \begin{subfigure}[b]{0.48\textwidth}
        \includegraphics[width=\textwidth]{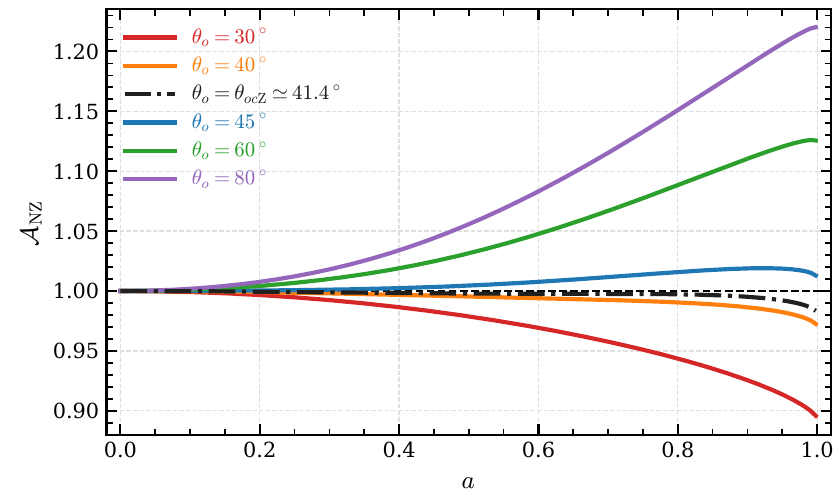}
        \caption{}
    \end{subfigure}
    \caption{The normalized shadow area of the rotating Ho\v{r}ava black hole at finite observer distances is plotted as a function of the spin parameter $a$ for different values of the LV factor $\zeta$ and the observer inclination angle $\theta_o$. The solid curves and open circles represent the approximate analytical and numerical results, respectively. Panel (a) illustrates the dependence on $a$ for different values of $\zeta$ at a fixed observer inclination angle $\theta_o = 80^\circ$, whereas panel (b) illustrates the corresponding dependence for different observer inclination angles $\theta_o$ at a fixed value of the LV factor $\zeta=4$.}
    \label{fig9}
\end{figure}

The dependence of the rotating Ho\v{r}ava black hole shadow area on the spin parameter $a$ at finite observer distances is illustrated in Fig. \ref{fig9} for different observer inclination angles and LV factors, with the results obtained using both approximate analytical and numerical methods.
The solid curves in Fig. \ref{fig9} represent the approximate analytical results, whereas the open circles in Fig. \ref{fig9} (a) denote the corresponding numerical results. The excellent agreement between the approximate analytical and numerical results demonstrates that the approximate analytical method provides a reliable alternative to numerical calculations for reproducing the shadow region of rotating Ho\v{r}ava black holes at finite observer distances.

The shadow area of rotating Ho\v{r}ava black holes is plotted as a function of the spin parameter $a$ in Fig. \ref{fig9} (a) for an observer located at $r_o = 6$ with an inclination angle $\theta_o = 80^\circ$ and LV factors $\zeta = 1$, $2$, $3$, and $4$. The $\zeta = 1$ configuration corresponds to the Kerr black hole, whose shadow area on the ZAMO image plane decreases monotonically with increasing $a$ and remains below the Schwarzschild value for nonzero spin, a behavior represented by the blue curve in Fig. \ref{fig9} (a). As the LV factor increases below the critical value $\zeta_{{c \text{Z}}} = 1.421$, the shadow area retains its monotonically decreasing trend, although the magnitude of the reduction progressively diminishes. When $\zeta$ exceeds $\zeta_{{c \text{Z}}}$, the shadow area instead increases monotonically with $a$ and remains above the Schwarzschild value for nonzero spin. The magnitude of this increase is progressively enhanced with the LV factor $\zeta$ and reaches a maximum at $\zeta = 4$ among the values considered. With the LV factor held fixed at $\zeta = 4$, the dependence of the rotating Ho\v{r}ava black hole shadow area on the spin parameter $a$ for different observer inclination angles $\theta_o$ is illustrated in Fig. \ref{fig9} (b). At $\theta_o = 30^\circ$, the shadow area decreases monotonically with increasing $a$ and remains below the Schwarzschild value for nonzero spin. As $\theta_o$ increases, the monotonic decrease persists, although the magnitude of the reduction becomes progressively smaller. When the observer inclination angle exceeds the critical value $\theta_{oc \text{Z}} = 41.4^\circ$, the shadow area exhibits a transition from a monotonic decrease to a monotonic increase with increasing $a$, rising above the Schwarzschild value. The increase in the shadow area becomes progressively more pronounced with increasing $\theta_o$.

The shadow area of rotating Ho\v{r}ava black holes at finite distances exhibits an additional dependence on the radial coordinate of the observer. Accordingly, the influence of the radial position of the observer on the anomalous increase in the shadow area warrants further investigation.
\begin{figure}[htbp]
    \centering
    \begin{subfigure}[b]{0.48\textwidth}
        \includegraphics[width=\textwidth]{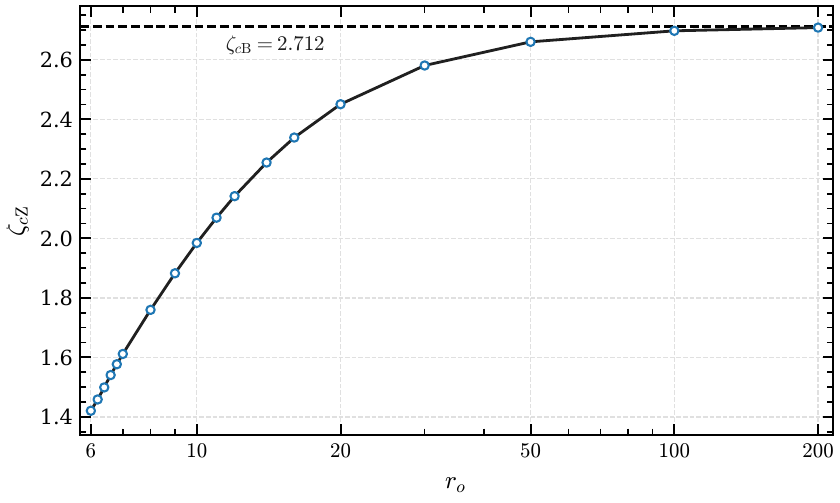}
        \caption{}
    \end{subfigure}
    \begin{subfigure}[b]{0.48\textwidth}
        \includegraphics[width=\textwidth]{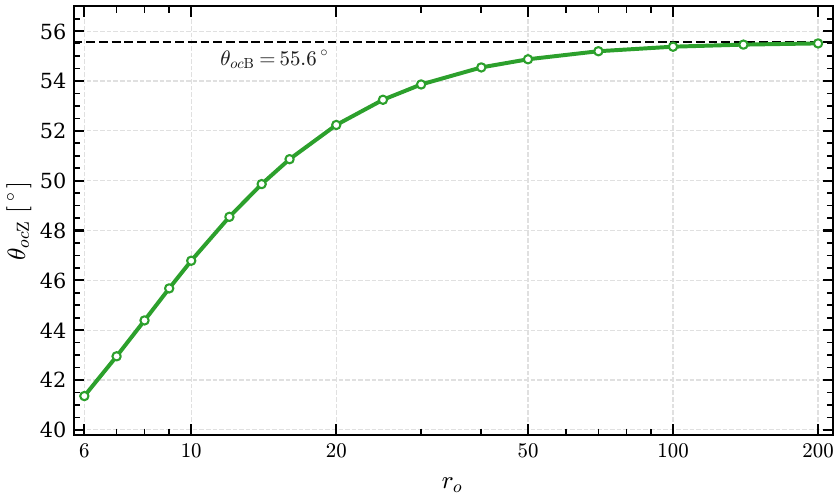}
        \caption{}
    \end{subfigure}
    \caption{The critical LV factor $\zeta_{c \text{Z}}$ and the critical observer inclination angle $\theta_{oc \text{Z}}$ are plotted as functions of the radial position $r_o$ of the observer. Panels (a) and (b) illustrate the dependence of $\zeta_{c \text{Z}}$ on $r_o$ at a fixed observer inclination angle $\theta_o = 80^\circ$ and that of $\theta_{oc \text{Z}}$ at a fixed value of the LV factor $\zeta = 4$, respectively.}
    \label{fig10}
\end{figure}
The critical LV factor $\zeta_{c \text{Z}}$ at $\theta_o = 80^\circ$ and the critical observer inclination angle $\theta_{oc \text{Z}}$ at $\zeta = 4$ are plotted as functions of the radial coordinate $r_o$ in Figs. \ref{fig10} (a) and (b), respectively. The critical LV factor and critical observer inclination angle initially increase approximately linearly with $r_o$ before eventually approaching the asymptotic values $\zeta_{{c \text{B}}} = 2.712$ and $\theta_{{oc \text{B}}} = 55.6^\circ$ as the observer approaches spatial infinity. The convergence of these critical values further supports the asymptotic correspondence between the black hole shadow configurations at finite observer distances and spatial infinity.

\subsubsection{Discussion of the anomalous increase in the shadow area at finite distances}

The anomalous increase in the rotating Ho\v{r}ava black hole shadow area persists for observers at finite distances, and the close agreement between the analytical and numerical results provides a basis for investigating the underlying mechanism analytically. However, the area integral derived from Green's theorem is more complicated at finite observer distances than at spatial infinity and cannot be evaluated to obtain an explicit closed-form expression for the shadow area. Consequently, in analogy with the analysis at spatial infinity, the finite-distance shadow area is analyzed through a series expansion in the spin parameter $a$.

The analytical treatment of the anomalous increase in the rotating Ho\v{r}ava black hole shadow area retains the Schwarzschild shadow area as the normalization reference to eliminate the influence of the scaling factor associated with the asymptotic relation in Eq. (\ref{finalrelationbardeenzamo}) and simplify the expression for the shadow area. Accordingly, the normalized shadow area can be expanded with respect to the spin parameter $a$ as
\begin{equation}\label{normalizedseriesexpressionshadowarea}
\frac{\mathcal A_{\rm Z}(a\,,\zeta)}{A_0} = 1 + \frac{c_{2\,, {\rm Z}} (\zeta)}{A_0} a^2 + \frac{c_{4\,, {\rm Z}} (\zeta)}{A_0} a^4 + \mathcal{O} (a^6)\,.
\end{equation}
The quantity $A_0$ denotes the Schwarzschild black hole shadow area at a finite observer distance and can be expressed explicitly as
\begin{equation}\label{seriesexpressiona0inzamo}
  A_0 = A_{Z,\,, \text{Sch}} = \frac{27 \pi \mathcal{L}^2 \Delta_{0}}{\left(r^2 + \sqrt{r^4 - 27 \Delta_{0}} \right)^2}\,,
\end{equation}
where $\Delta_{0} = \left. \Delta_r \right|_{a=0} = r (r - 2)$. The asymptotic relation in Eq. (\ref{finalrelationbardeenzamo}) further allows the expression for $A_0$ in the limit $r \rightarrow \infty$ to be written as
\begin{equation}
  \begin{split}
    \lim_{r \to \infty} \left( \frac{2 r}{\mathcal{L}} \right)^2 A_0 = 27 \pi\,.
  \end{split}
\end{equation}
The preceding asymptotic relation indicates that the black hole shadow image constructed at finite observer distances approaches the corresponding shadow image at spatial infinity as the radial coordinate of the observer increases.

The higher-order coefficients in Eq. (\ref{normalizedseriesexpressionshadowarea}) can, in principle, be determined from the area integral derived from Green's theorem. However, the expression for the second-order coefficient at finite observer distances is more complicated than the expression obtained at spatial infinity, and an exact closed-form expression for this coefficient cannot be obtained by straightforward analytical evaluation. To simplify the analytical treatment of the shadow area, the second-order coefficient is further expanded in the asymptotic limit $r\rightarrow\infty$, and the resulting expression is given by
\begin{equation}\label{seriesexpressionc2inzamo}
  \frac{c_{2\,, \text{Z}} (\zeta)}{A_0} = - \frac{1 + \cos^2 \theta_O}{18} + \frac{16 \sin^2 \theta_o \zeta - 5 (1 + \cos^2 \theta_o)}{4 r^2} + \mathcal{O} (r^{-3})\,.
\end{equation}
It should be emphasized that this asymptotic expansion is not intended merely to approximate the area of the rotating Ho\v{r}ava black hole shadow image constructed in the ZAMO frame at spatial infinity. The asymptotic result is instead employed as a reference for identifying the corrections at finite observer distances associated with the radial displacement of the observer from spatial infinity toward the black hole. The leading term in Eq. (\ref{seriesexpressionc2inzamo}) agrees exactly with the second-order coefficient in Eq. (\ref{coeffinseriesexpan}) for the black hole shadow area on the Bardeen image plane at spatial infinity. At finite observer distances, however, the second-order coefficient in Eq. (\ref{normalizedseriesexpressionshadowarea}) depends on the observer inclination angle $\theta_o$, the radial coordinate $r_o$, and the LV factor $\zeta$. The cancellation of contributions involving the LV factor to the second-order coefficient at spatial infinity no longer holds as the radial coordinate $r_o$ decreases, and the second-order coefficient consequently acquires a nontrivial dependence on $\zeta$. The calculation of the fourth-order coefficient in Eq. (\ref{normalizedseriesexpressionshadowarea}) can likewise be formulated in terms of the area integral derived from Green's theorem, although the substantially greater algebraic complexity relative to the second-order case prevents the derivation of an exact closed-form expression by direct integration. The following analysis therefore examines the properties of the fourth-order coefficient graphically without presenting its explicit analytical expression.

\begin{figure}[t]
  \centering
  \includegraphics[width=0.6\linewidth]{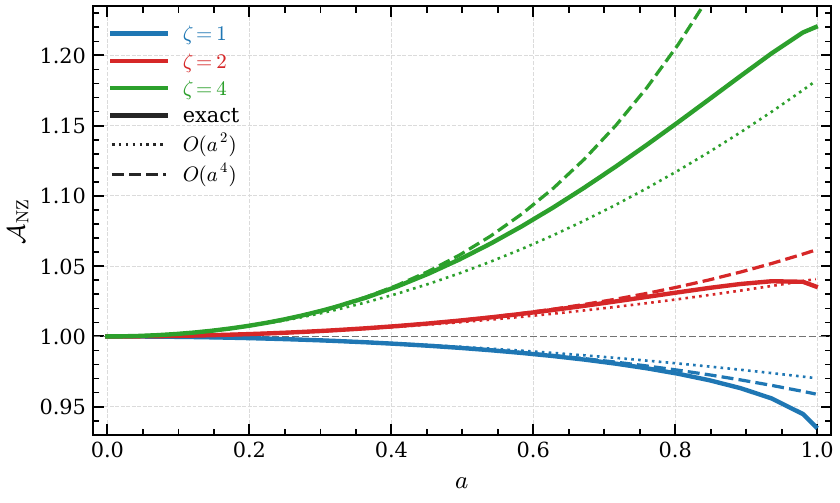}
  \caption{The rotating Ho\v{r}ava black hole shadow area at finite observer distances is calculated as a function of the spin parameter $a$ within the approximate analytical framework for $\zeta=1$, $2$, and $4$ at a fixed observer inclination angle $\theta_o = 80^\circ$. The solid curves represent the full analytical expression for the black hole shadow area, whereas the dotted and dashed curves are obtained from the series expansions truncated at second- and fourth-order with respect to the spin parameter, respectively.}
  \label{fig11}
\end{figure}
Fig. \ref{fig11} compares the variation of the rotating Ho\v{r}ava black hole shadow area with the spin parameter $a$ as calculated from the full analytical expression and the series expansions truncated at second and fourth order for an observer located at $r_o = 6$ with an inclination angle $\theta_o = 80^\circ$ and LV factors $\zeta = 1$, $2$, and $4$. The solid, dotted, and dashed curves represent the shadow areas calculated from the full analytical expression and the series expansions truncated at second and fourth order, respectively. The horizontal dashed line denotes the Schwarzschild black hole shadow area. In the static limit $a = 0$, the full analytical result and the second- and fourth-order truncated expansions reproduce the Schwarzschild shadow area independently of the LV factor $\zeta$, in agreement with the leading term of Eq. (\ref{normalizedseriesexpressionshadowarea}) and the Schwarzschild limit of the rotating Ho\v{r}ava solution. In the Kerr limit $\zeta = 1$, the shadow area decreases monotonically with increasing $a$. Within the interval $a\in[0,\, 0.6]$, the second-order term provides the principal contribution to the reduction in the shadow area. For $a\in[0.6,\, 0.8]$, the second-order approximation no longer adequately captures the continued decrease, while the fourth-order contribution becomes dominant and accounts for a further reduction in the shadow area. With the LV factor $\zeta = 2$ exceeding the critical value $\zeta_{{c \text{Z}}}$, the shadow area instead exhibits an anomalous monotonic increase with $a$. The increase in the shadow area is dominated by the second-order term over the interval $a\in[0,\, 0.6]$ and by the fourth-order term over the interval $a\in[0.6,\, 0.8]$. At $\zeta = 4$, the increase in the shadow area is primarily governed by the second-order term for $a\in[0,\, 0.2]$, with the fourth-order term providing an increasingly dominant contribution within the range $a\in[0.2,\, 0.4]$. These results demonstrate that the second-order contribution already plays a dominant role in the anomalous increase in the rotating Ho\v{r}ava black hole shadow area at finite observer distances. Furthermore, as the LV factor $\zeta$ increases, the second- and fourth-order terms contribute appreciably at progressively smaller values of the spin parameter $a$.

The dominant role of the second-order term in the anomalous increase in the rotating Ho\v{r}ava black hole shadow area at finite observer distances motivates a more detailed investigation of the associated expansion coefficient. Eq. (\ref{seriesexpressionc2inzamo}) reveals that the second-order coefficient depends explicitly on the radial coordinate $r_o$ of the observer and the LV factor $\zeta$. The full analytical expression for the second-order coefficient is therefore employed to quantitatively characterize this dependence without relying on the truncated asymptotic series expansion. With the observer inclination angle fixed at $\theta_o = 80^\circ$, the coefficient values obtained for different combinations of $r_o$ and $\zeta$ are summarized in Table \ref{table3}.
\begin{table}[t]
  \caption{The first three columns list the values of the second-order coefficient $c_{2,\, \text{Z}} (\zeta) / A_0$ for different values of the LV factor $\zeta$ at various radial positions $r_o$ of the observer. The last column reports the corresponding relative deviation $\Delta_{41}$ as a function of $r_o$.}
  \centering
  \begin{tabular}{ccccc}
    \hline
    $r_O$ & $\zeta=1$ & $\zeta=2$ & $\zeta=4$ & $\Delta_{41}$ \\
    \hline
    $6$   & $-0.0298$  & $+0.0410$  & $+0.1826$  & $0.2124$\\
    $10$  & $-0.0428$  & $-0.0148$  & $+0.0414$ & $0.0842$\\
    $30$  & $-0.0549$  & $-0.0512$  & $-0.0437$ & $0.0112$\\
    $200$ & $-0.05717$ & $-0.05708$ & $-0.05692$ & $2.5 \times 10^{-4}$\\
    \hline
  \end{tabular}
  \label{table3}
\end{table}
The results in Table \ref{table3} show that the second-order coefficient at $r_o=6$ increases with the LV factor $\zeta$ and undergoes a transition from negative to positive values. The second-order term therefore contributes to a reduction in the shadow area at small LV factors and to an increase once $\zeta$ exceeds the critical value $\zeta_{c \text{Z}}$, with the largest positive contribution attained at $\zeta = 4$ among the values considered. With the radial coordinate increased to $r_o = 10$, the second-order coefficient continues to increase with $\zeta$, attaining a positive value only at $\zeta = 4$ among the LV factors considered. The positive coefficient is smaller than that obtained at $r_o = 6$ for the same LV factor, implying a weaker enhancement of the shadow area at second order. At the larger radial coordinate $r_o = 30$, the coefficient continues to increase with $\zeta$ while remaining negative throughout the parameter range considered, consistently contributing to a reduction in the shadow area. For a distant observer located at $r_o = 200$, the second-order coefficient remains negative and is nearly independent of the LV factor $\zeta$.

The rotating Ho\v{r}ava black hole solution reduces to the Kerr solution at $\zeta = 1$, providing a reference for assessing the influence of the LV factor on the second-order coefficient in the shadow area expansion. The quantity $\Delta_{41}$ is introduced as a quantitative measure of the deviation of the second-order coefficient at $\zeta = 4$ from the Kerr coefficient at $\zeta = 1$ and is defined as
\begin{equation}
  \Delta_{41} \equiv \left(\frac{c_{2,{\rm Z}}}{A_0}\right)_{\zeta=4} - \left(\frac{c_{2,{\rm Z}}}{A_0}\right)_{\zeta=1}.
\end{equation}
The last column of Table \ref{table3} presents the values of the deviation $\Delta_{41}$ at different radial coordinates $r_o$. The deviation $\Delta_{41}$ exhibits a monotonic decrease with increasing $r_o$ and attains a nearly vanishing value at $r_o = 200$. This radial dependence indicates that the influence of the LV factor on the second-order coefficient is more pronounced for observers in the vicinity of the black hole. The deviation from the Kerr coefficient progressively diminishes with increasing observer distance and is negligible at sufficiently large radial coordinates. In the limit $r_o \rightarrow \infty$, the rotating Ho\v{r}ava black hole shadow area therefore agrees with the Kerr shadow area up to and including second order in the spin parameter $a$.

Furthermore, the second-order truncation of the rotating Ho\v{r}ava black hole shadow area retaining the full analytical expression for the second-order expansion coefficient is denoted by $\mathcal{A}_{\text{NZ}}^{(2)}$ and employed to characterize the dependence of the shadow area on the spin parameter $a$ for different combinations of $\zeta$ and $r_o$.
\begin{figure}[t]
  \centering
  \includegraphics[width=0.95\linewidth]{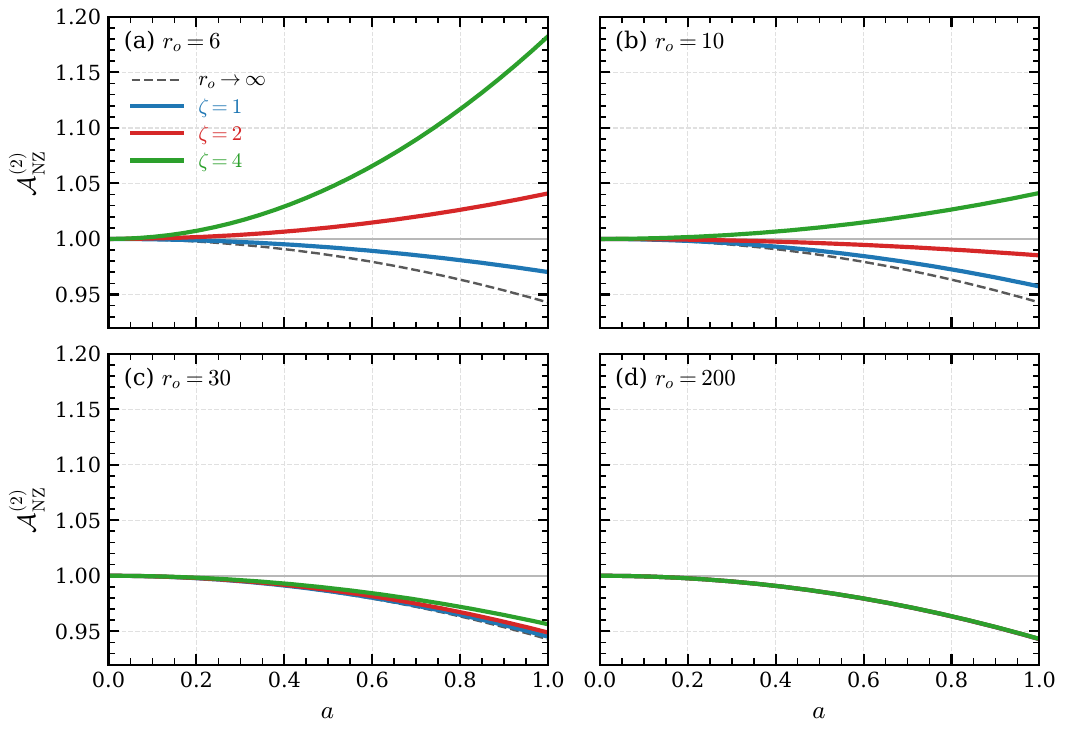}
  \caption{The second-order truncation $\mathcal{A}_{\text{NZ}}^{(2)}$ of the full analytical expression for the rotating Ho\v{r}ava black hole shadow area at finite observer distances is plotted as a function of the spin parameter $a$. The solid curves represent the second-order approximations to the shadow area for $\zeta=1,2$ and $4$, whereas the black dashed curve denotes the corresponding second-order approximation at spatial infinity. Panels (a), (b), (c), and (d) display the results for observer radial positions $r_o=6,10,30$ and $200$, respectively.}
  \label{fig12}
\end{figure}
Figs. \ref{fig12} (a), (b), (c), and (d) present the rotating Ho\v{r}ava black hole shadow area calculated from $\mathcal{A}_{\text{NZ}}^{(2)}$ as a function of the spin parameter $a$ for observers with a fixed inclination angle $\theta_o = 80^\circ$ and radial coordinates $r_o = 6$, $10$, $30$, and $200$, respectively. The blue, red, and green curves correspond to the LV factors $\zeta=1$, $2$, and $4$, respectively, while the black dashed curve represents the second-order approximation to the Kerr black hole shadow area at spatial infinity. In Fig. \ref{fig12} (a), the blue curve describes the monotonic decrease in the Kerr black hole shadow area with increasing $a$ for an observer located at $r_o=6$. The discrepancy between the blue and black dashed curves indicates that the shadow area depends on the radial coordinate of the observer even for the Kerr spacetime. For LV factors exceeding the critical value, the second-order term induces an anomalous increase in the shadow area with $a$ that is progressively more pronounced at larger values of $\zeta$. Figs. \ref{fig12} (b), (c), and (d) further reveal that the enhancement of the shadow area induced by the second-order contribution progressively diminishes with increasing observer distance, accompanied by the gradual convergence of the curves toward the black dashed curve. At $r_o = 200$, the curves are visually indistinguishable from the second-order approximation to the Kerr black hole shadow area at spatial infinity.

The convergence of these curves with increasing $r_o$ demonstrates that the second-order coefficient in the shadow area expansion of the rotating Ho\v{r}ava black hole progressively approaches that of the Kerr black hole at spatial infinity. The asymptotic coincidence of the curves for different LV factors is consistent with the exact cancellation of contributions involving $\zeta$ established analytically at spatial infinity. The explicit dependence of the second-order coefficient on $\zeta$ at finite observer distances therefore reflects the absence of this exact cancellation. The progressive restoration of this cancellation as the observer approaches spatial infinity accounts for the recovery of a second-order coefficient independent of $\zeta$ and identical to that of the Kerr black hole in the asymptotic limit.

\subsubsection{Source of LV factor in the second-order term at finite distances}

The LV factor $\zeta$ enters explicitly into the expressions for both the photon impact parameters and the associated projection coordinates on the ZAMO image plane, as given in Eqs. (\ref{tretaalambdaexpanda}) and (\ref{specificexpzamocoordinates}), respectively. The contributions involving $\zeta$ to the second-order term in the shadow area expansion of the rotating Ho\v{r}ava black hole may therefore originate from photon dynamics as well as from the projection of photon trajectories onto the image plane. To identify the physical origin of these contributions, five auxiliary parameters $p_i,\, (i =1,\, 2,\, 3,\, 4,\, 5)$ are introduced to distinguish the individual terms involving the LV factor in the photon impact parameters and the image plane projection coordinates.

The photon impact parameter $\eta$ in Eq. (\ref{tretaalambdaexpanda}) is reexpressed in terms of the parameters $p_1$ and $p_2$ as
\begin{equation}\label{impactparameteretawithp}
  \eta = 27 - \lambda^2 - 4 p_1 \sqrt{\zeta} \lambda a + \left(\frac{5}{27} \lambda^2 - 4 - p_2 \frac{8 \zeta}{27} \lambda^2 \right) a^2 + \mathcal{O} (a^3)\,.
\end{equation}
The parameters $p_1$ and $p_2$ both characterize contributions involving the LV factor that originate from photon dynamics. In the projection coordinates, the explicit dependence on the LV factor enters through the function $\mathcal{D} (\lambda,, \eta)$. The introduction of the parameters $p_3$, $p_4$, and $p_5$ enables this function to be expressed in parametrized form as
\begin{equation}\label{functionDwithparameterp}
  \begin{split}
    \mathcal{D} (\lambda\,, \eta) = \Sigma^2 - 2 p_3 a r \sqrt{\zeta} \lambda + \sqrt{\Sigma^2 \mathcal{H} (\lambda\,, \eta)}\,,
  \end{split}
\end{equation}
where the auxiliary function $\mathcal{H} (\lambda,, \eta)$ is given explicitly as
\begin{equation}\label{functionHwithparameterp}
  \begin{split}
    \mathcal{H} (\lambda\,, \eta) = r^2 (r^2 + a^2) + 2 r a^2 - \Delta \eta - 4 p_4 a r \sqrt{\zeta} \lambda + r (2 - r) \lambda^2 + p_5 \frac{4 (\zeta - 1) a^2 r^2}{\Sigma^2} \lambda^2\,.
  \end{split}
\end{equation}
The parameters $p_3$, $p_4$, and $p_5$ accordingly characterize the contributions involving the LV factor that enter the rotating Ho\v{r}ava black hole shadow area through the expressions for the photon projection coordinates on the ZAMO image plane.

The normalized second-order coefficient in the shadow area expansion of the rotating Ho\v{r}ava black hole is recalculated from Eqs. (\ref{impactparameteretawithp}), (\ref{functionDwithparameterp}), and (\ref{functionHwithparameterp}) by applying the same analytical procedure and expanded asymptotically as
\begin{equation}\label{secondordertermwithparameterp}
  \begin{split}
    \frac{c_{2,{\text{Z}}} (\zeta)}{A_0} = - \frac{1 + \cos^2 \theta_O}{18} + \frac{16 \sin^2 \theta_o (3 p_1^2 - p_2) \zeta - 5 (1 + \cos^2 \theta_o)}{4 r^2} + \mathcal{O} (r^{-3})\,.
  \end{split}
\end{equation}
The dependence of the second-order coefficient in Eq. (\ref{secondordertermwithparameterp}) on the auxiliary parameters is restricted to $p_1$ and $p_2$, whereas the coefficient is independent of $p_3$, $p_4$, and $p_5$ within the present asymptotic approximation. The occurrence of $p_1$ and $p_2$ exclusively in the expression for the photon impact parameter $\eta$ indicates that the LV dependence of the second-order coefficient within this approximation arises solely from photon dynamics and is unaffected by the terms explicitly involving $\zeta$ in the projection coordinates on the ZAMO image plane. With the auxiliary parameters fixed at $p_1 = p_2 = 1$, Eq. (\ref{secondordertermwithparameterp}) reduces to the asymptotic expression in Eq. (\ref{seriesexpressionc2inzamo}).

\section{Conclusions}\label{section5}

The spacetime geometry of rotating Ho\v{r}ava black holes is generally of Petrov type I. The Hamilton-Jacobi equation governing photon motion in these spacetimes does not admit an exact separation of variables, which precludes the direct analytical construction of black hole shadows by conventional separation methods. The non-separable contribution to the Hamilton-Jacobi equation is identified as the function $T (r,\, \theta)$, which exhibits a predominantly radial dependence and only a weak angular dependence. This property permits an approximate separation of the Hamilton-Jacobi equation, providing a basis for the analytical construction of rotating Ho\v{r}ava black hole shadows. The approximate analytical method is employed to construct shadow images for observers at spatial infinity and at finite radial distances, and the analytical predictions are independently verified through the numerical backward ray tracing method. The excellent agreement between the analytical and numerical results establishes the reliability of the approximate analytical approach as a complementary method for investigating rotating Ho\v{r}ava black hole shadows. Furthermore, an asymptotic relation between the projection schemes employed in constructing black hole shadow images at finite observer distances and at spatial infinity is established through the scaling factor $2 r / \mathcal{L}$.

The analytical investigation is subsequently extended to the shadow area of the rotating Ho\v{r}ava black hole at spatial infinity. The shadow area is found to exhibit an anomalous increase with the spin parameter $a$ beyond the critical values identified for the observer inclination angle and the LV factor. The physical mechanism underlying this anomalous behavior is elucidated by expanding the shadow area in powers of $a$. The second-order coefficient coincides with that of the Kerr black hole as a consequence of the exact cancellation of all contributions involving the LV factor, while the fourth-order coefficient is the first to exhibit an explicit dependence on the LV factor. The comparison between the second- and fourth-order truncated expansions establishes the predominant role of the fourth-order contribution in the anomalous increase in the shadow area. Furthermore, as the LV factor increases, the fourth-order term produces a larger positive correction to the shadow area and contributes significantly at progressively smaller values of the spin parameter $a$.

The anomalous increase in the shadow area of the rotating Ho\v{r}ava black hole with the spin parameter $a$ persists at finite observer distances when the observer inclination angle and the LV factor exceed the respective critical values. As the radial coordinate of the observer increases, these critical values progressively approach those obtained at spatial infinity. The series expansion reveals that the explicit dependence of the second-order coefficient on $\zeta$ at finite observer distances arises from the absence of the exact cancellation of contributions involving the LV factor established at spatial infinity. The comparison between the second- and fourth-order truncated expansions demonstrates that the anomalous increase in the shadow area is governed primarily by the second-order term, whereas the fourth-order term provides an additional enhancement at larger values of the spin parameter $a$. The enhancement attributable to the second-order term progressively diminishes with increasing observer distance, and the second-order coefficient asymptotically approaches that of the Kerr black hole at spatial infinity. This convergence reflects the progressive restoration of the cancellation among contributions involving the LV factor, which is exact in the asymptotic limit. Finally, the dependence of the second-order coefficient on the LV factor $\zeta$ within the asymptotic approximation considered is attributable solely to photon dynamics rather than to the explicit dependence on the LV factor in the photon projection coordinates on the image plane.

\section*{Acknowledgments}
This work is partially supported by the National Natural Science Foundation of China (NSFC) under Grant No. 12575048 and No. 12105015. X. -Y. Wang is also supported by the Guangdong Basic and Applied Basic Research Foundation under Grant No. 2023A1515012737. M. Guo is also supported by the BNU Tang Scholar.

\appendix

\section*{Appendix}

\section{Numerical construction of rotating Ho\v{r}ava black hole shadow images at spatial infinity}\label{appendixa}

\begin{figure}[htbp]
    \centering
    \includegraphics[width=0.5\textwidth]{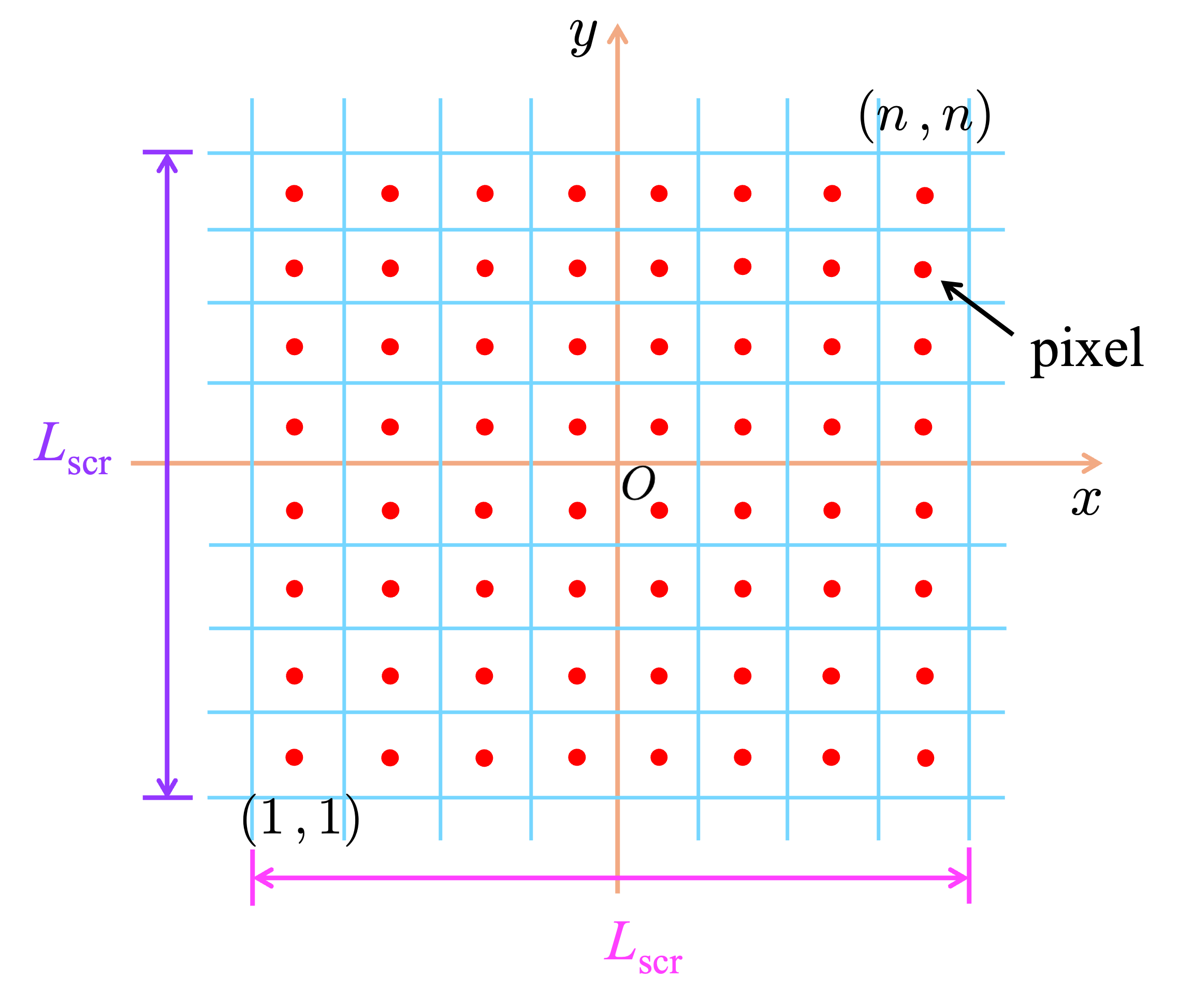}
    \caption{The discretization of the image plane is illustrated schematically. The $x$ and $y$ axes define a Cartesian coordinate system on the image plane. The solid blue lines delineate the boundaries of the grid cells, whereas the red dots indicate the pixel positions corresponding to the projection of photons onto the image plane. The parameter $L_{\text{scr}}$ specifies both the length and the width of the square imaging region.}
    \label{figA1}
\end{figure}

The numerical construction of the rotating Ho\v{r}ava black hole shadow at spatial infinity through backward ray tracing is initiated by the discretization of the image plane, as schematically illustrated in Fig. \ref{figA1}.
The $x$ and $y$ axes in Fig. \ref{figA1} define the Cartesian coordinate system on the image plane, while the blue solid lines delineate the discretized grid. The red point represents the pixel associated with the projection of a photon reaching the observer onto the image plane. The image plane is chosen to be a square of side length $L_{\text{scr}}$ and is uniformly discretized into an $n \times n$ grid, with each grid cell representing a single pixel. These $n^2$ pixels are indexed by $(i,\, j)$ to specify the horizontal and vertical positions on the image plane. The indices satisfy $1 \leq i,\, j \leq n$, with $(1,\, 1)$ and $(n,\, n)$ assigned to the pixels at the lower left and upper right corners, respectively. The uniform spacing between adjacent pixel centers along the $x$ and $y$ directions is given by
\begin{equation}
  \begin{split}
     \Delta_{\text{pix}} = \frac{L_{\text{scr}}}{n}\,.
  \end{split}
\end{equation}
Therefore, the coordinates of the pixels on the discretized image plane can be written as
\begin{equation}\label{gridpixelcoordinates}
  \begin{split}
    x_{ij} = \Delta_{\text{pix}} \left(i - \frac{n + 1}{2} \right)\,, \qquad y_{ij} = \Delta_{\text{pix}} \left(j - \frac{n + 1}{2}\right)\,.
  \end{split}
\end{equation}
The pixel coordinates on the image plane of an observer at spatial infinity are directly identified with the Bardeen coordinates through the relation $\left(x_{ij},\, y_{ij} \right)=\left(x_B,\, y_B \right)$.

The initial conditions for the numerical construction of the rotating Ho\v{r}ava black hole shadow through backward ray tracing are specified by the spacetime position of the observer $x_o^a=(t_o,\, r_o,\, \theta_o,\, \phi_o)$ and the covariant components of the photon four-momentum $p_{a o}$ evaluated at the observer position. In the numerical implementation, the observational configuration at spatial infinity is approximated by setting the radial coordinate of the observer to a value representative of the asymptotic region, $r_o = 1000$. For simplicity, the temporal and azimuthal coordinates are fixed at $t_o = \phi_o = 0$, whereas the observer inclination angle $\theta_o$ is specified prior to each numerical calculation. The spacetime position of the observer is therefore expressed as $x_o^a = (0,\, 1000,\, \theta_o,\, 0)$.

The initial covariant components of the photon four-momentum are specified as $p_{a o} =(p_{t o},\, p_{r o},\, p_{\theta o},\, p_{\phi o})$. The definitions of the conserved photon energy and angular momentum determine the temporal and azimuthal components as $p_{t o} = - E$ and $p_{\phi o} = L$, respectively. With the photon energy normalized to $E = 1$, the temporal component reduces to $p_{t o} = -1$. For the adopted energy normalization, the azimuthal component is obtained from Eqs. (\ref{defimpactparameters}) and (\ref{bardeencoordinate}) as $p_{\phi o} = - x_B \sin \theta_o$. The polar component of the photon four-momentum follows from the $\theta$ component of the equations of motion in Eq. (\ref{equationofmotionbl}) and the Bardeen coordinate relation in Eq. (\ref{bardeencoordinate}) as $p_{\theta o} = y_B$. The radial component of the initial photon four-momentum is determined by imposing the null condition $p_a p^a=0$ as
\begin{equation}
  \begin{split}
    p_{r o} = - \sqrt{- \frac{g^{t t} p_{t o}^2 + 2 g^{t \phi} p_{t o} p_{\phi o} + g^{\phi \phi} p_{\phi o}^2 + g^{\theta \theta} p_{\theta o}^2}{g^{r r}}}\,,
  \end{split}
\end{equation}
where the negative sign is selected to ensure that the photon trajectory is integrated from the observer toward the black hole in the numerical backward ray tracing method. The initial covariant components of the photon four-momentum can accordingly be expressed as $p_{a o} = (-1\,, p_{r o}\,, y_B\,, - x_B \sin \theta_o)$.

The numerical construction of rotating Ho\v{r}ava black hole shadows through backward ray tracing is based on the canonical Hamilton equations governing photon geodesic motion in the black hole spacetime. These equations can be expressed as
\begin{equation}\label{hamiltoniancanonicalequations}
  \begin{split}
    \dot{x}_a = \frac{\partial H}{\partial p_a}\,, \qquad \dot{p}_a = - \frac{\partial H}{\partial x^a}\,,
  \end{split}
\end{equation}
where the Hamiltonian $H$ is defined in Eq. (\ref{definitionhamiltonian}). The photon trajectory in the rotating Ho\v{r}ava spacetime is obtained by integrating the canonical Hamilton equations subject to the initial conditions specified by the observer position $x_o^a$ and the photon four-momentum $p_{a o}$. In the numerical backward ray tracing method, the rotating Ho\v{r}ava black hole shadow image is constructed in Bardeen coordinates by integrating the photon trajectory associated with each pixel on the image plane and determining whether the photon is captured by the black hole or escapes to spatial infinity.

\section{Analytical and numerical construction of rotating Ho\v{r}ava black hole shadows at finite distance}

\subsection{Introduction of the ZAMO frame and asymptotic relation to the Bardeen coordinates}\label{zamobardeenrelation}

\begin{figure}[htbp]
    \centering
    \begin{subfigure}[b]{0.4\textwidth}
        \includegraphics[width=\textwidth]{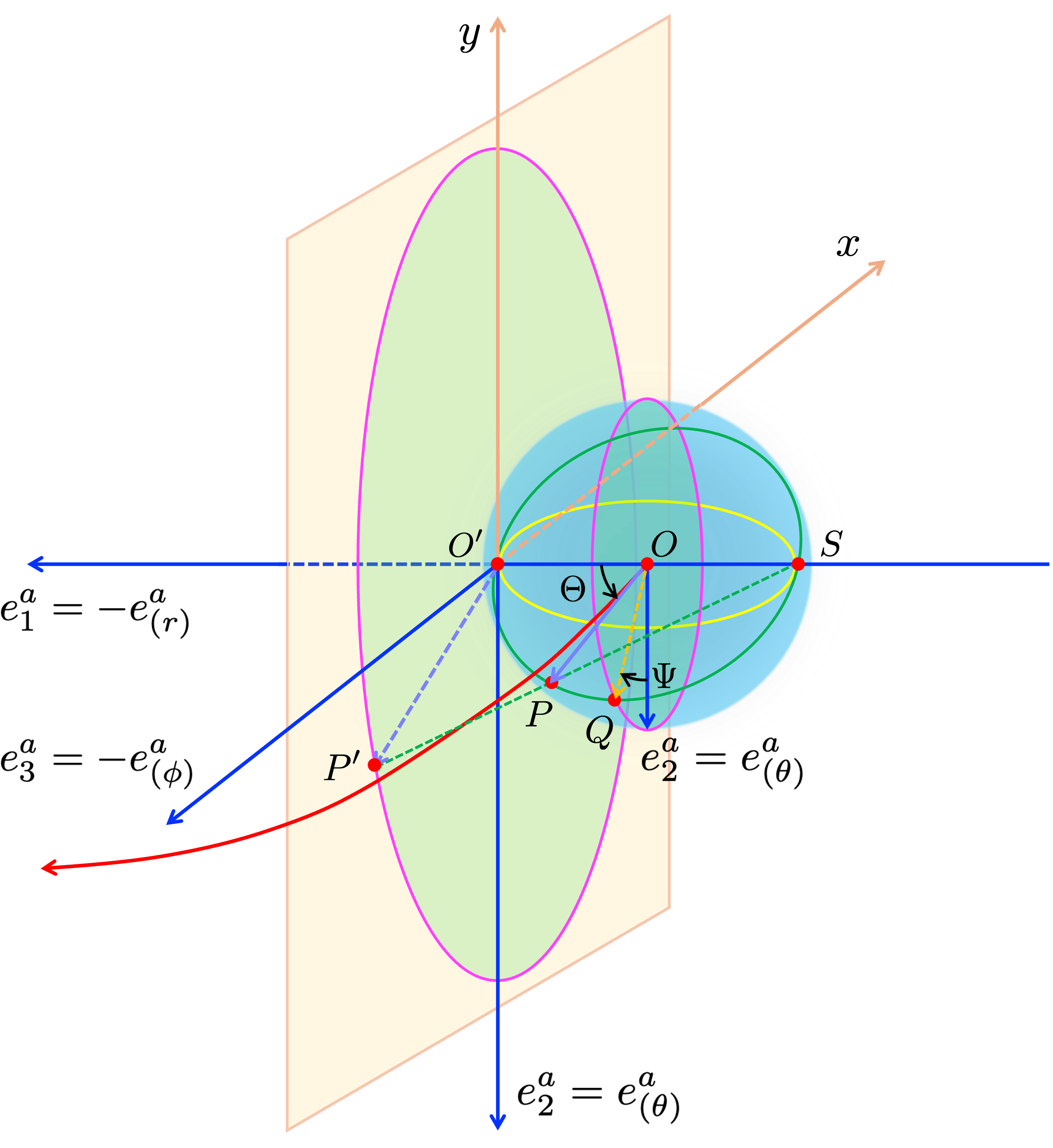}
        \caption{}
    \end{subfigure}
    \begin{subfigure}[b]{0.38\textwidth}
        \includegraphics[width=\textwidth]{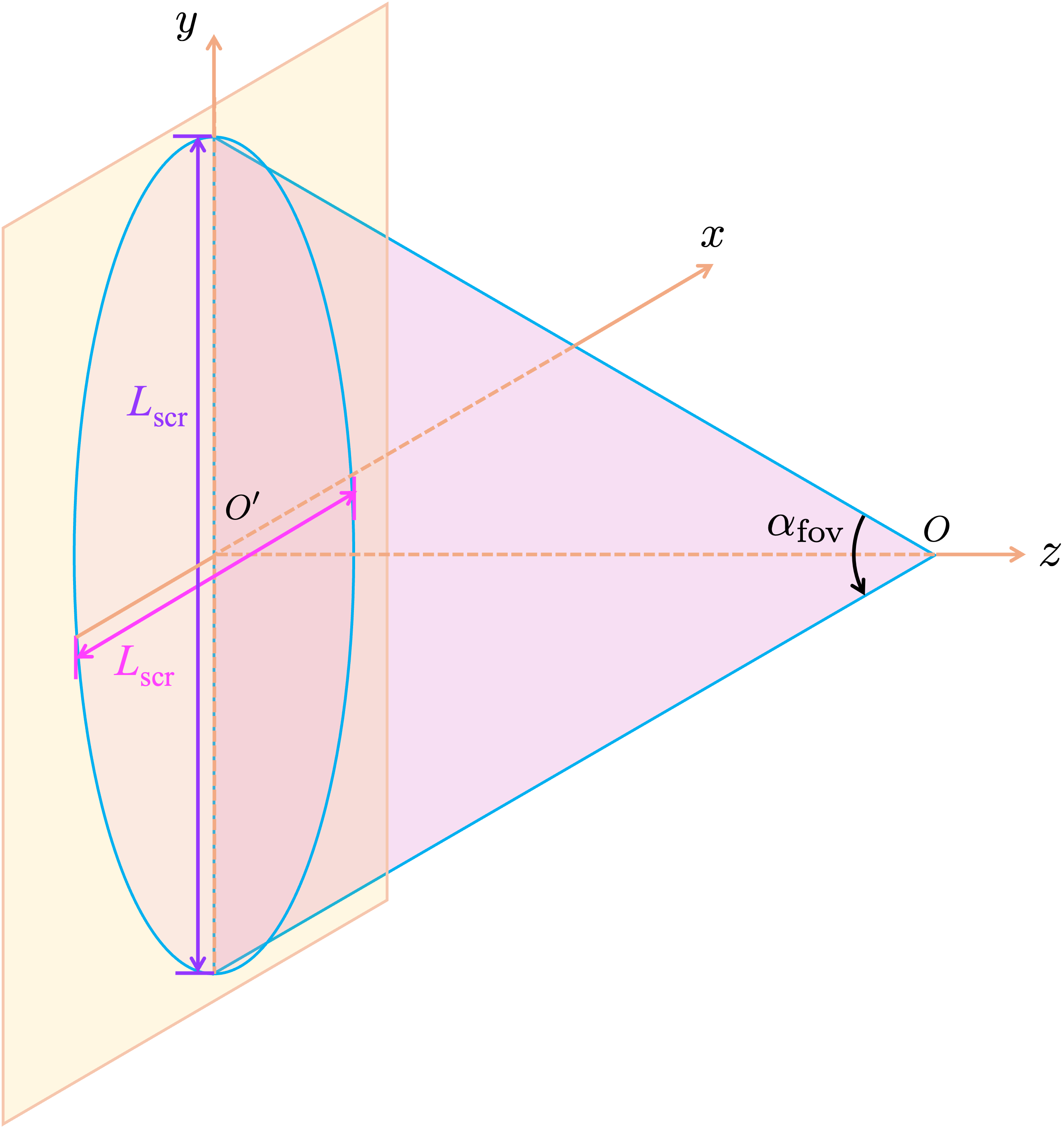}
        \caption{}
    \end{subfigure}
    \caption{Panel (a) schematically illustrates the stereographic projection of photon propagation directions measured in the ZAMO frame onto the image plane. Panel (b) depicts the maximum angular extent of the field of view $\alpha_{\text{fov}}$ and the corresponding imaging region on the image plane \cite{Wang:2025ihg}}.
    \label{figB1}
\end{figure}

The construction of black hole shadow images at finite observer distances is formulated in a local orthonormal frame associated with a ZAMO to characterize photon propagation as measured by the observer. The basis vectors of the ZAMO frame are defined as
\begin{equation}\label{definitionzamobasis}
  \begin{split}
  e_{(t)}^a & = \sqrt{\frac{g_{\phi \phi}}{g_{t \phi}^2 - g_{tt} g_{\phi \phi}}} \left[\left(\partial_t \right)^a - \frac{g_{t \phi}}{g_{\phi \phi}} \left(\partial_\phi \right)^a \right]\,,\\
  e_{(r)}^a & = - \frac{1}{\sqrt{g_{r r}}} \left(\partial_r \right)^a\,,\\
  e_{(\theta)}^a & = \frac{1}{\sqrt{g_{\theta \theta}}} \left(\partial_\theta \right)^a\,,\\
  e_{(\phi)}^a & = - \frac{1}{\sqrt{g_{\phi \phi}}} \left(\partial_\phi \right)^a\,.
  \end{split}
\end{equation}
The black hole shadow image in the ZAMO frame is constructed by stereographically projecting the local photon propagation directions onto the image plane, as schematically illustrated by the projection geometry in Fig. \ref{figB1}.

Fig. \ref{figB1} (a) illustrates the geometrical configuration of the ZAMO frame and the associated image plane for an observer located at a finite distance from the black hole. The blue solid lines denote the spatial axes of the ZAMO frame, while the orange plane represents the image plane. The observer is located at point $O$, and the red solid curve and the associated arrow indicate the photon trajectory and the direction of photon propagation, respectively. The vector $\overrightarrow{OP}$ represents the spatial three-momentum obtained by projecting the photon four-momentum onto the three-dimensional spatial subspace orthogonal to the four-velocity of the ZAMO and is tangent to the spatial projection of the photon trajectory at $O$. To determine the projection coordinates on the image plane, two celestial angles are introduced to parameterize the photon propagation direction measured in the ZAMO frame. As illustrated in Fig. \ref{figB1} (a), the angle between the photon three-momentum $\overrightarrow{OP}$ and the spatial basis vector $e_1$ is denoted by $\Theta$. The vector $\overrightarrow{OQ}$ represents the orthogonal projection of $\overrightarrow{OP}$ onto the plane passing through $O$ and perpendicular to $e_1$, while $\Psi$ denotes the angle between $\overrightarrow{OQ}$ and the spatial basis vector $e_2$. The celestial coordinates in the ZAMO frame are specified by the angles $(\Theta,\Psi)$. On the basis of the coordinate transformation presented in Ref. \cite{Hu:2020usx, Wang:2025ihg}, the stereographic projection coordinates on the ZAMO image plane can be expressed in terms of $(\Theta,\, \Psi)$ as
\begin{equation}\label{stereographicprojection}
  \begin{split}
    x_{\text{Z}} = -\mathcal L \tan \frac{\Theta}{2} \sin \Psi\,, \qquad y_{\text{Z}} = -\mathcal L \tan \frac{\Theta}{2} \cos \Psi\,,
  \end{split}
\end{equation}
where $\mathcal{L}$ is a characteristic scale parameter of the stereographic camera model. The stereographic projection of directions parameterized by the celestial coordinates onto the ZAMO image plane illustrated in Fig. \ref{figB1} (a) corresponds to a fisheye camera model. The viewing angle $\alpha_{\text{fov}}$ quantifies the camera field of view in terms of the maximum angular extent of the imaging region on the image plane. For simplicity, the imaging domain is restricted to a square of side length $L_{\text{scr}}$ centered at the Cartesian origin of the image plane and bounded by lines parallel to the $x$ and $y$ axes, as illustrated in Fig. \ref{figB1} (b). The relation between the camera parameter $\mathcal{L}$ and the maximum viewing angle $\alpha_{\text{fov}}$ can be expressed in terms of the side length $L_{\text{scr}}$ as
\begin{equation}
  \mathcal{L} = L_{\text{scr}} \left(\tan \frac{\alpha_{\rm fov}}{2} \right)^{-1}\,.
\end{equation}

The photon four-momentum can be expressed in terms of the celestial coordinates and the orthonormal basis vectors of the ZAMO frame as
\begin{equation}\label{fourmomentumphotonzamo}
  \begin{split}
    p^{a} = - e_{(t)}^a + \cos \Theta e_{(r)}^a + \sin \Theta \cos \Psi e_{(\theta)}^a + \sin \Theta \sin \Psi e_{(\phi)}^a\,,
  \end{split}
\end{equation}
where the energy parameter associated with the photon four-momentum directed toward the past in the ZAMO frame is normalized to $E = - p_a e_{(t)}^a = - 1$. The energy normalization can be imposed without loss of generality, as any nonzero constant rescaling of the photon four-momentum can be absorbed into a redefinition of the affine parameter without modifying the spacetime trajectory of the null geodesic. Furthermore, the photon projection coordinates on the ZAMO image plane given in Eq. (\ref{stereographicprojection}) can be recast by applying the null condition $p_a p^a=0$ in conjunction with the relevant trigonometric identities as
\begin{equation}\label{xzyzsinthetasinpsicospsi}
  \begin{split}
    x_{\text{Z}} = & - \mathcal{L} \frac{\sin \Theta \sin \Psi}{1 + \sqrt{1 - (\sin \Theta \sin \Psi)^2 - (\sin \Theta \cos \Psi)^2}},\\
    y_{\text{Z}} = & - \mathcal{L} \frac{\sin \Theta \cos \Psi}{1 + \sqrt{1 - (\sin \Theta \sin \Psi)^2 - (\sin \Theta \cos \Psi)^2}}.
  \end{split}
\end{equation}
The factors $\sin \Theta \sin \Psi$ and $\sin \Theta \cos \Psi$ in Eq. (\ref{xzyzsinthetasinpsicospsi}) represent the $\phi$ and $\theta$ components of the photon four-momentum in Eq. (\ref{fourmomentumphotonzamo}), respectively. The azimuthal component is derived by combining the energy normalization in the ZAMO frame with the relation between the conserved photon energy and angular momentum established by the definition of the impact parameter $\lambda$ in Eq. (\ref{impactparameters}). The polar component is obtained by relating the $\theta$ component of the photon four-momentum in Eq. (\ref{fourmomentumphotonzamo}) to the corresponding component of the photon equations of motion in Eq. (\ref{equationofmotionbl}). The resulting expressions for these two factors are given by
\begin{equation}\label{sinThetasinPsi}
  \begin{split}
    \sin \Theta \sin \Psi = \frac{\lambda \sqrt{g_{t \phi}^{2} - g_{t t} g_{\phi \phi}}}{g_{\phi \phi} + g_{t \phi} \lambda}\,, \qquad 
    \sin \Theta \cos \Psi = \sigma_\theta E \sqrt{\frac{\Theta (\theta_o)}{g_{\theta \theta}}}\,.
  \end{split}
\end{equation}

The asymptotic analysis of the rotating Ho\v{r}ava spacetime in the limit $r \rightarrow \infty$ provides a basis for establishing the correspondence between the photon projection coordinates on the image planes at finite observer distances and at spatial infinity. The metric components of the rotating Ho\v{r}ava spacetime in Eq. (\ref{rotatinghoravametric}) can be expanded asymptotically as
\begin{equation}\label{asymptoticmetriccomponents}
  \begin{split}
    g_{tt} = & - 1 + \frac{2}{r} - \frac{2 \left(a^2 \cos^2 \theta \right)}{r^3} + \mathcal{O} (r^{- 4})\,,\\
    g_{r r} = & 1 + \frac{2}{r} + \frac{4 - a^2 + a^2 \cos^2 \theta}{r^2} + \mathcal{O} (r^{- 3})\,,\\ 
    g_{\theta \theta} = & r^2 + a^2 \cos^2 \theta + \mathcal{O} (r^{- 1})\,,\\
    g_{\phi \phi} = & r^2 \sin^2 \theta \left(1 + \frac{a^2}{r^2} + \frac{2 a^2 \sin^2 \theta}{r^3} + \mathcal{O} (r^{-4}) \right)\,,\\
    g_{t \phi} = & - \frac{2 a \sqrt{\zeta} \sin^2 \theta}{r} + \frac{2 a^3 \sqrt{\zeta} \sin^2 \theta \cos^2 \theta}{r^3} + \mathcal{O} (r^{- 4})\,.
  \end{split}
\end{equation}
With the metric components replaced by the asymptotic expansions in Eq. (\ref{asymptoticmetriccomponents}), the asymptotic forms of the ZAMO basis vectors defined in Eq. (\ref{definitionzamobasis}) are obtained as
\begin{equation}
  \begin{split}
    e_{(t)} = & \left(1 + \frac{1}{r} + \frac{3}{2 r^2} + \mathcal{O} (r^{- 3}) \right) \left[\partial_t + \left(\frac{2 a \sqrt{\zeta}}{r^3} + \mathcal{O} (r^{-5}) \right) \partial_\phi \right]\,,\\
    e_{(r)} = & \left(-1 + \frac{1}{r} + \frac{1 - a^2 + a^2 \cos^2 \theta}{2 r^2} + \mathcal{O} (r^{-3}) \right) \partial_r\,,\\
    e_{(\theta)} = & \left(\frac{1}{r} - \frac{a^2 \cos^2 \theta}{2 r^3} + \mathcal{O} (r^{- 4})\right) \partial_\theta\,,\\
    e_{(\phi)} = & \left(- \frac{1}{r \sin \theta} + \frac{a^2}{2 r^3 \sin \theta} + \frac{a^2 \sin \theta}{r^4} + \mathcal{O} (r^{- 5})\right) \partial_\phi\,.
  \end{split}
\end{equation}
The asymptotic expansions of the ZAMO basis vectors reveal that the ZAMO frame progressively approaches that of a static observer in the limit $r \rightarrow \infty$. The LV factor $\zeta$ first enters the asymptotic expansion of the ZAMO frame through the coefficient of $\partial_\phi$ in the temporal basis vector $e_{(t)}$. Since the ratio of the coefficient of $\partial_\phi$ to that of $\partial_t$ in $e_{(t)}$ defines the coordinate angular velocity of the ZAMO, the leading dependence of the ZAMO frame on the LV factor originates from the modification of the angular velocity induced by frame dragging. The corresponding angular velocity at the position of the observer can be expressed as
\begin{equation}
  \Omega_{\text{ZAMO}} \equiv - \frac{g_{t \phi}}{g_{\phi \phi}} = \frac{2 a \sqrt{\zeta}}{r^3} + \mathcal O(r^{-5})\,.
\end{equation}

The metric expansions in Eq. (\ref{asymptoticmetriccomponents}) are applied to Eq. (\ref{sinThetasinPsi}) to express the $\phi$ and $\theta$ components of the photon four-momentum given in Eq. (\ref{fourmomentumphotonzamo}) asymptotically as
\begin{equation}\label{sinThetasinPsisinThetacosPsi}
  \begin{split}
    \sin \Theta \sin \Psi = \frac{\lambda}{r \sin \theta_o} + \mathcal{O} (r^{-2})\,, \qquad \sin \Theta \cos \Psi = - \frac{\sigma_{\theta} \sqrt{\Theta (\theta_o)}}{r} + \mathcal{O} (r^{-2})\,.
  \end{split}
\end{equation}
The asymptotic expressions in Eq. (\ref{sinThetasinPsisinThetacosPsi}) can be reformulated in terms of the Bardeen coordinates defined by Eq. (\ref{bardeencoordinate}) as
\begin{equation}\label{sinThetasinPsisinThetacosPsixbyb}
  \begin{split}
    \left(\sin \Theta \sin \Psi,\, \sin \Theta \cos \Psi \right) = - \frac{1}{r} \left(x_{\text{B}},\, y_{\text{B}} \right)
  +\mathcal O (r^{-2}).
  \end{split}
\end{equation}
The pixel coordinates on the ZAMO image plane given in Eq. (\ref{stereographicprojection}) can be expressed in an equivalent form through the relevant trigonometric identities as
\begin{equation}\label{zamocoordinateswithouttan}
  \begin{split}
    \left(x_{\text{Z}},\, y_{\text{Z}}\right) = - \frac{\mathcal{L}}{1 + \cos \Theta} \left(\sin \Theta \sin \Psi\,, \sin \Theta \cos \Psi \right)\,.
  \end{split}
\end{equation}
The asymptotic radial behavior of the two factors in Eq. (\ref{sinThetasinPsisinThetacosPsi}) is characterized by $\sin \Theta \sin \Psi \sim \mathcal{O} (r^{-1})$ and $\sin \Theta \cos \Psi \sim \mathcal{O} (r^{-1})$. The radial scaling relations for the two factors together with the approximation valid for small values of $\Theta$ establish $\Theta^2 \sim \sin^2 \Theta \sim \mathcal{O} (r^{-2})$ and consequently $\Theta \sim \mathcal{O} (r^{-1})$ as $r \rightarrow \infty$. The factor $(1 + \cos \Theta)^{-1}$ in Eq. (\ref{zamocoordinateswithouttan}) can therefore be expanded asymptotically as
\begin{equation}\label{inverse1pluscosTheta}
  \begin{split}
    (1 + \cos \Theta)^{-1} = \frac{1}{2} + \frac{\Theta^2}{8} + \mathcal{O} (\Theta^{4}) = \frac{1}{2} + \mathcal{O} (r^{-2})\,.
  \end{split}
\end{equation}
The asymptotic relation between the photon projection coordinates on the ZAMO and Bardeen image planes can be derived from Eqs. (\ref{sinThetasinPsisinThetacosPsixbyb}), (\ref{zamocoordinateswithouttan}), and (\ref{inverse1pluscosTheta}) as
\begin{equation}\label{coordinatesrelationbardeenzamo}
  \begin{split}
    \left(x_{\text{Z}},\, y_{\text{Z}}\right) = \frac {\mathcal{L}}{2 r} \left(x_{\text{B}},\, y_{\text{B}} \right) + \mathcal{O} (r^{-2})\,.
  \end{split}
\end{equation}
The result in Eq. (\ref{coordinatesrelationbardeenzamo}) demonstrates that the leading terms in the asymptotic expansions of the ZAMO image plane coordinates differ from the corresponding Bardeen coordinates at spatial infinity only by a common scaling factor of $\mathcal{L} / (2 r)$.

\subsection{Construction of black hole shadow in ZAMO coordinates by approximate analytical and numerical method}\label{zamoanalyticalnumericalmethod}

The general expressions for the photon projection coordinates on the ZAMO image plane are derived by combining Eqs. (\ref{sinThetasinPsi}) and (\ref{xzyzsinthetasinpsicospsi}) with the metric components of the rotating Ho\v{r}ava spacetime specified in Eq. (\ref{rotatinghoravametric}) as
\begin{equation}\label{specificexpzamocoordinates}
  \begin{split}
    x_{\text{Z}}(\lambda,\, \eta) = -\mathcal L \frac{\rho^2 \lambda \sqrt{\Delta}}{\mathcal{D} (\lambda,\, \eta) \sin \theta_o}\,, \qquad y_{\text{Z}}(\lambda,\, \eta;\, \sigma) = \sigma_{\theta} \mathcal{L} \frac{\sqrt{\Sigma^2 \Delta \Theta (\theta_o)}}{\mathcal{D} (\lambda,\, \eta)}\,,
  \end{split}
\end{equation}
where the function $\mathcal{D} (\lambda,\, \eta)$ is defined as
\begin{equation}
  \begin{split}
    \mathcal{D} (\lambda,\, \eta) = \Sigma^2 - 2 a r \sqrt{\zeta} \lambda + \sqrt{\Sigma^2 \mathcal{H} (\lambda,\, \eta)}
  \end{split}
\end{equation}
and the auxiliary function $\mathcal{H} (\lambda,\, \eta)$ entering $\mathcal{D} (\lambda,\, \eta)$ is given explicitly as
\begin{equation}
  \begin{split}
    \mathcal{H} (\lambda,\, \eta) = r^2 (r^2 + a^2) + 2 r a^2 - \Delta \eta - 4 a r \sqrt{\zeta} \lambda + r (2 - r) \lambda^2 + \frac{4 (\zeta - 1) a^2 r^2}{\Sigma^2} \lambda^2\,.
  \end{split}
\end{equation}
The mapping from the photon impact parameters to the projection coordinates on the ZAMO image plane established in Eq. (\ref{specificexpzamocoordinates}) provides the mathematical basis for the approximate analytical construction of rotating Ho\v{r}ava black hole shadow images at finite observer distances.

For the numerical construction of the black hole shadow image at finite observer distances, the ZAMO image plane is discretized using the same scheme employed at spatial infinity and illustrated in Fig. \ref{figA1}. The computational grid comprises $n^2$ cells obtained by uniformly subdividing the image plane into $n$ intervals along each Cartesian direction. The center of each grid cell specifies the position of the corresponding pixel indexed by $(i,\, j)$ with $1 \leq i,\, j \leq n$. The Cartesian coordinates of an arbitrary pixel on the image plane are specified by Eq. (\ref{gridpixelcoordinates}). The celestial coordinates associated with the pixel indices $(i,\, j)$ are determined by imposing equality between the photon projection coordinates in Eq. (\ref{stereographicprojection}) and the pixel coordinates defined in Eq. (\ref{gridpixelcoordinates}). The celestial coordinates $(\Theta,\, \Psi)$ can accordingly be expressed as
\begin{equation}
  \begin{split}
    \tan\frac{\Theta}{2} & = \frac{1}{n} \tan \left(\frac{\alpha_{\text{fov}}}{2} \right) \sqrt{\left(i - \frac{n + 1}{2} \right)^2 + \left(j - \frac{n + 1}{2} \right)^2}\,,\\
    \tan\Psi & = \frac{i - \frac{n + 1}{2}}{j - \frac{n + 1}{2}}\,.
  \end{split}
\end{equation}
The geometrical relations $\sin \Psi = - x_{ij} / \rho$ and $\cos \Psi = - y_{ij} / \rho$ specify the celestial angle $\Psi$ for any pixel at a nonzero radial distance $\rho = \sqrt{x_{ij}^2 + y_{ij}^2}$ from the center of the ZAMO image plane. The expressions for the photon projection coordinates in Eq. (\ref{stereographicprojection}) can be combined with these geometrical relations and the definition of $\rho$ to express the celestial coordinates $(\Theta,\, \Psi)$ in terms of the pixel coordinates $(x_{ij},\, y_{ij})$ as
\begin{equation}\label{celestialcoordinates}
  \begin{split}
    c_{(r)} & \equiv \cos \Theta = \frac{\mathcal L^2 - \rho^2}{\mathcal L^2 + \rho^2}\,,\\
    c_{(\theta)} & \equiv \sin \Theta \cos \Psi = - \frac{2 \mathcal L  y_{ij}}{\mathcal L^2 + \rho^2}\,,\\
    c_{(\phi)} & \equiv \sin \Theta \sin \Psi = - \frac{2 \mathcal L x_{ij}}{\mathcal L^2 + \rho^2}\,,
  \end{split}
\end{equation}
The three direction cosines in the ZAMO frame are subject to the normalization constraint $c_{(r)}^2 + c_{(\theta)}^2 + c_{(\phi)}^2 = 1$. The initial covariant components of the photon four-momentum are derived from Eqs. (\ref{celestialcoordinates}) and (\ref{fourmomentumphotonzamo}) as
\begin{equation}\label{initialcovmomentumzamoframe}
  \begin{split}
    p_{t o} & = \sqrt{\frac{g_{t \phi}^2 - g_{tt} g_{\phi \phi}}{g_{\phi \phi}}} - \frac{g_{t \phi}}{\sqrt{g_{\phi \phi}}} c_{(\phi)}\,,\\
    p_{r o} & = - \sqrt{g_{r r}} c_{(r)}\,,\\
    p_{\theta o} & = \sqrt{g_{\theta \theta}} c_{(\theta)}\,,\\
    p_{\phi o} & = - \sqrt{g_{\phi \phi}} c_{(\phi)}\,.
  \end{split}
\end{equation}
The expressions in Eq. (\ref{initialcovmomentumzamoframe}) for the covariant components of the photon four-momentum specify the initial momenta required for the numerical construction of the rotating Ho\v{r}ava black hole shadow image through backward ray tracing. To complete the initial conditions for the numerical integration, the spacetime position of the observer is parameterized as $(t_o,\, r_o,\, \theta_o,\, \phi_o)=(0,\, r_o,\, \theta_o,\, 0)$. The radial coordinate $r_o$ and the observer inclination angle $\theta_o$ are prescribed according to the observational configuration considered in the numerical analysis. The camera parameter is fixed at $\mathcal{L} = 2$ in accordance with Eq. (\ref{finalrelationbardeenzamo}) to ensure the asymptotic convergence of the ZAMO shadow image rescaled by $r_o$ to the Bardeen shadow image at spatial infinity. The numerical integration of the canonical Hamilton equations in Eq. (\ref{hamiltoniancanonicalequations}) with the prescribed spacetime position of the observer and the covariant components of the photon four-momentum as initial conditions determines the photon trajectory associated with an individual pixel on the image plane. The rotating Ho\v{r}ava black hole shadow image on the ZAMO image plane is constructed by numerically integrating the photon trajectories associated with all pixels and classifying each pixel according to whether the corresponding photon is captured by the black hole or escapes to spatial infinity.

\bibliographystyle{utphys}

\bibliography{References}

@article{EventHorizonTelescope:2019dse,
    author = "Akiyama, Kazunori and others",
    collaboration = "Event Horizon Telescope",
    title = "{First M87 Event Horizon Telescope Results. I. The Shadow of the Supermassive Black Hole}",
    eprint = "1906.11238",
    archivePrefix = "arXiv",
    primaryClass = "astro-ph.GA",
    doi = "10.3847/2041-8213/ab0ec7",
    journal = "Astrophys. J. Lett.",
    volume = "875",
    pages = "L1",
    year = "2019"
}

@article{EventHorizonTelescope:2022xqj,
    author = "Akiyama, Kazunori and others",
    collaboration = "Event Horizon Telescope",
    title = "{First Sagittarius A* Event Horizon Telescope Results. VI. Testing the Black Hole Metric}",
    eprint = "2311.09484",
    archivePrefix = "arXiv",
    primaryClass = "astro-ph.HE",
    reportNumber = "FERMILAB-PUB-22-422-PPD",
    doi = "10.3847/2041-8213/ac6756",
    journal = "Astrophys. J. Lett.",
    volume = "930",
    number = "2",
    pages = "L17",
    year = "2022"
}

@article{Perlick:2021aok,
    author = "Perlick, Volker and Tsupko, Oleg Yu.",
    title = "{Calculating black hole shadows: Review of analytical studies}",
    eprint = "2105.07101",
    archivePrefix = "arXiv",
    primaryClass = "gr-qc",
    doi = "10.1016/j.physrep.2021.10.004",
    journal = "Phys. Rept.",
    volume = "947",
    pages = "1--39",
    year = "2022"
}

@article{Grenzebach:2014fha,
    author = {Grenzebach, Arne and Perlick, Volker and L{\"a}mmerzahl, Claus},
    title = "{Photon Regions and Shadows of Kerr-Newman-NUT Black Holes with a Cosmological Constant}",
    eprint = "1403.5234",
    archivePrefix = "arXiv",
    primaryClass = "gr-qc",
    doi = "10.1103/PhysRevD.89.124004",
    journal = "Phys. Rev. D",
    volume = "89",
    number = "12",
    pages = "124004",
    year = "2014"
}

@article{Gralla:2019xty,
    author = "Gralla, Samuel E. and Holz, Daniel E. and Wald, Robert M.",
    title = "{Black Hole Shadows, Photon Rings, and Lensing Rings}",
    eprint = "1906.00873",
    archivePrefix = "arXiv",
    primaryClass = "astro-ph.HE",
    doi = "10.1103/PhysRevD.100.024018",
    journal = "Phys. Rev. D",
    volume = "100",
    number = "2",
    pages = "024018",
    year = "2019"
}

@article{Bronzwaer:2021lzo,
    author = "Bronzwaer, Thomas and Falcke, Heino",
    title = "{The Nature of Black Hole Shadows}",
    eprint = "2108.03966",
    archivePrefix = "arXiv",
    primaryClass = "astro-ph.HE",
    doi = "10.3847/1538-4357/ac1738",
    journal = "Astrophys. J.",
    volume = "920",
    number = "2",
    pages = "155",
    year = "2021"
}

@article{Hioki:2009na,
    author = "Hioki, Kenta and Maeda, Kei-ichi",
    title = "{Measurement of the Kerr Spin Parameter by Observation of a Compact Object's Shadow}",
    eprint = "0904.3575",
    archivePrefix = "arXiv",
    primaryClass = "astro-ph.HE",
    reportNumber = "WU-AP-299-09",
    doi = "10.1103/PhysRevD.80.024042",
    journal = "Phys. Rev. D",
    volume = "80",
    pages = "024042",
    year = "2009"
}

@article{Zhong:2021mty,
    author = "Zhong, Zhen and Hu, Zezhou and Yan, Haopeng and Guo, Minyong and Chen, Bin",
    title = "{QED effects on Kerr black hole shadows immersed in uniform magnetic fields}",
    eprint = "2108.06140",
    archivePrefix = "arXiv",
    primaryClass = "gr-qc",
    doi = "10.1103/PhysRevD.104.104028",
    journal = "Phys. Rev. D",
    volume = "104",
    number = "10",
    pages = "104028",
    year = "2021"
}

@article{Cunha:2019ikd,
    author = "Cunha, Pedro V. P. and Herdeiro, Carlos A. R. and Radu, Eugen",
    title = "{EHT constraint on the ultralight scalar hair of the M87 supermassive black hole}",
    eprint = "1909.08039",
    archivePrefix = "arXiv",
    primaryClass = "gr-qc",
    doi = "10.3390/universe5120220",
    journal = "Universe",
    volume = "5",
    number = "12",
    pages = "220",
    year = "2019"
}

@article{Kumar:2020owy,
    author = "Kumar, Rahul and Ghosh, Sushant G.",
    title = "{Rotating black holes in $4D$ Einstein-Gauss-Bonnet gravity and its shadow}",
    eprint = "2003.08927",
    archivePrefix = "arXiv",
    primaryClass = "gr-qc",
    doi = "10.1088/1475-7516/2020/07/053",
    journal = "JCAP",
    volume = "07",
    pages = "053",
    year = "2020"
}

@article{Synge:1966okc,
    author = "Synge, J. L.",
    title = "{The Escape of Photons from Gravitationally Intense Stars}",
    doi = "10.1093/mnras/131.3.463",
    journal = "Mon. Not. Roy. Astron. Soc.",
    volume = "131",
    number = "3",
    pages = "463--466",
    year = "1966"
}

@article{Teo:2003ltt,
    author = "Teo, Edward",
    title = "{Spherical Photon Orbits Around a Kerr Black Hole}",
    doi = "10.1023/A:1026286607562",
    journal = "Gen. Rel. Grav.",
    volume = "35",
    number = "11",
    pages = "1909--1926",
    year = "2003"
}

@article{deVries:1999tiy,
    author = "de Vries, A.",
    title = "{The apparent shape of a rotating charged black hole, closed photon orbits and the bifurcation set $A_4$}",
    doi = "10.1088/0264-9381/17/1/309",
    journal = "Class. Quant. Grav.",
    volume = "17",
    number = "1",
    pages = "123--144",
    year = "1999"
}

@article{Feng:2019zzn,
    author = "Feng, Xing-Hui and Lu, H.",
    title = "{On the size of rotating black holes}",
    eprint = "1911.12368",
    archivePrefix = "arXiv",
    primaryClass = "gr-qc",
    doi = "10.1140/epjc/s10052-020-8119-z",
    journal = "Eur. Phys. J. C",
    volume = "80",
    number = "6",
    pages = "551",
    year = "2020"
}

@article{Cunha:2016bpi,
    author = "Cunha, Pedro V. P. and Herdeiro, Carlos A. R. and Radu, Eugen and Runarsson, Helgi F.",
    editor = "Herdeiro, Carlos A. R. and Berti, Emanuele and Cardoso, Vitor and Crispino, Luis C. B. and Gualtieri, Leonardo and Sperhake, Ulrich",
    title = "{Shadows of Kerr black holes with and without scalar hair}",
    eprint = "1605.08293",
    archivePrefix = "arXiv",
    primaryClass = "gr-qc",
    doi = "10.1142/S0218271816410212",
    journal = "Int. J. Mod. Phys. D",
    volume = "25",
    number = "09",
    pages = "1641021",
    year = "2016"
}

@article{Younsi:2016azx,
    author = "Younsi, Ziri and Zhidenko, Alexander and Rezzolla, Luciano and Konoplya, Roman and Mizuno, Yosuke",
    title = "{New method for shadow calculations: Application to parametrized axisymmetric black holes}",
    eprint = "1607.05767",
    archivePrefix = "arXiv",
    primaryClass = "gr-qc",
    doi = "10.1103/PhysRevD.94.084025",
    journal = "Phys. Rev. D",
    volume = "94",
    number = "8",
    pages = "084025",
    year = "2016"
}

@article{Bambi:2015kza,
    author = "Bambi, Cosimo",
    title = "{Testing black hole candidates with electromagnetic radiation}",
    eprint = "1509.03884",
    archivePrefix = "arXiv",
    primaryClass = "gr-qc",
    doi = "10.1103/RevModPhys.89.025001",
    journal = "Rev. Mod. Phys.",
    volume = "89",
    number = "2",
    pages = "025001",
    year = "2017"
}

@article{Johannsen:2010ru,
    author = "Johannsen, Tim and Psaltis, Dimitrios",
    title = "{Testing the No-Hair Theorem with Observations in the Electromagnetic Spectrum: II. Black-Hole Images}",
    eprint = "1005.1931",
    archivePrefix = "arXiv",
    primaryClass = "astro-ph.HE",
    doi = "10.1088/0004-637X/718/1/446",
    journal = "Astrophys. J.",
    volume = "718",
    pages = "446--454",
    year = "2010"
}

@article{Mizuno:2018lxz,
    author = "Mizuno, Yosuke and Younsi, Ziri and Fromm, Christian M. and Porth, Oliver and De Laurentis, Mariafelicia and Olivares, Hector and Falcke, Heino and Kramer, Michael and Rezzolla, Luciano",
    title = "{The Current Ability to Test Theories of Gravity with Black Hole Shadows}",
    eprint = "1804.05812",
    archivePrefix = "arXiv",
    primaryClass = "astro-ph.GA",
    doi = "10.1038/s41550-018-0449-5",
    journal = "Nature Astron.",
    volume = "2",
    number = "7",
    pages = "585--590",
    year = "2018"
}

@article{EventHorizonTelescope:2020qrl,
    author = "Psaltis, Dimitrios and others",
    collaboration = "Event Horizon Telescope",
    title = "{Gravitational Test Beyond the First Post-Newtonian Order with the Shadow of the M87 Black Hole}",
    eprint = "2010.01055",
    archivePrefix = "arXiv",
    primaryClass = "gr-qc",
    doi = "10.1103/PhysRevLett.125.141104",
    journal = "Phys. Rev. Lett.",
    volume = "125",
    number = "14",
    pages = "141104",
    year = "2020"
}

@article{Horava:2009uw,
    author = "Horava, Petr",
    title = "{Quantum Gravity at a Lifshitz Point}",
    eprint = "0901.3775",
    archivePrefix = "arXiv",
    primaryClass = "hep-th",
    doi = "10.1103/PhysRevD.79.084008",
    journal = "Phys. Rev. D",
    volume = "79",
    pages = "084008",
    year = "2009"
}

@article{Sotiriou:2010wn,
    author = "Sotiriou, Thomas P.",
    editor = "Perivolaropoulos, Leandros and Kanti, Panagiota",
    title = "{Horava-Lifshitz gravity: a status report}",
    eprint = "1010.3218",
    archivePrefix = "arXiv",
    primaryClass = "hep-th",
    doi = "10.1088/1742-6596/283/1/012034",
    journal = "J. Phys. Conf. Ser.",
    volume = "283",
    pages = "012034",
    year = "2011"
}

@article{Wang:2017brl,
    author = "Wang, Anzhong",
    title = "{Ho{\v{r}}ava gravity at a Lifshitz point: A progress report}",
    eprint = "1701.06087",
    archivePrefix = "arXiv",
    primaryClass = "gr-qc",
    doi = "10.1142/S0218271817300142",
    journal = "Int. J. Mod. Phys. D",
    volume = "26",
    number = "07",
    pages = "1730014",
    year = "2017"
}

@article{Blas:2009qj,
    author = "Blas, D. and Pujolas, O. and Sibiryakov, S.",
    title = "{Consistent Extension of Horava Gravity}",
    eprint = "0909.3525",
    archivePrefix = "arXiv",
    primaryClass = "hep-th",
    reportNumber = "CERN-PH-TH-2009-173",
    doi = "10.1103/PhysRevLett.104.181302",
    journal = "Phys. Rev. Lett.",
    volume = "104",
    pages = "181302",
    year = "2010"
}

@article{Barvinsky:2015kil,
    author = "Barvinsky, Andrei O. and Blas, Diego and Herrero-Valea, Mario and Sibiryakov, Sergey M. and Steinwachs, Christian F.",
    title = "{Renormalization of Ho{\v{r}}ava gravity}",
    eprint = "1512.02250",
    archivePrefix = "arXiv",
    primaryClass = "hep-th",
    reportNumber = "CERN-PH-TH-2015-287, INR-TH-2015-032, IFT-UAM-CSIC-15-127, FTUAM-15-44, FR-PHENO-2015-015",
    doi = "10.1103/PhysRevD.93.064022",
    journal = "Phys. Rev. D",
    volume = "93",
    number = "6",
    pages = "064022",
    year = "2016"
}

@article{Barausse:2011pu,
    author = "Barausse, Enrico and Jacobson, Ted and Sotiriou, Thomas P.",
    title = "{Black holes in Einstein-aether and Horava-Lifshitz gravity}",
    eprint = "1104.2889",
    archivePrefix = "arXiv",
    primaryClass = "gr-qc",
    doi = "10.1103/PhysRevD.83.124043",
    journal = "Phys. Rev. D",
    volume = "83",
    pages = "124043",
    year = "2011"
}

@article{Jacobson:2010mx,
    author = "Jacobson, Ted",
    title = "{Extended Horava gravity and Einstein-aether theory}",
    eprint = "1001.4823",
    archivePrefix = "arXiv",
    primaryClass = "hep-th",
    doi = "10.1103/PhysRevD.81.101502",
    journal = "Phys. Rev. D",
    volume = "81",
    pages = "101502",
    year = "2010",
    note = "[Erratum: Phys.Rev.D 82, 129901 (2010)]"
}

@article{Devecioglu:2024Rotating,
    author = "Devecio\u{g}lu, Deniz O. and Park, Mu-In",
    title = "{Rotating Black Holes in a Viable Lorentz-Violating Gravity: Finding Exact Solutions Without Tears}",
    eprint = "2402.02253",
    archivePrefix = "arXiv",
    primaryClass = "hep-th",
    doi = "10.1140/epjc/s10052-024-13209-3",
    journal = "Eur. Phys. J. C",
    volume = "84",
    number = "8",
    pages = "852",
    year = "2024"
}

@article{Medeiros:2020NonKerr,
    author = "Medeiros, Lia and Psaltis, Dimitrios and {\"O}zel, Feryal",
    title = "{A Parametric Model for the Shapes of Black Hole Shadows in Non-Kerr Spacetimes}",
    eprint = "1907.12575",
    archivePrefix = "arXiv",
    primaryClass = "astro-ph.HE",
    doi = "10.3847/1538-4357/ab8bd1",
    journal = "Astrophys. J.",
    volume = "896",
    number = "1",
    pages = "7",
    year = "2020"
}

@article{Psaltis:2021Metric,
    author = "Psaltis, Dimitrios and Talbot, Colm and Payne, Ethan and Mandel, Ilya",
    title = "{Probing the Black Hole Metric. I. Black Hole Shadows and Binary Black-Hole Inspirals}",
    eprint = "2012.02117",
    archivePrefix = "arXiv",
    primaryClass = "gr-qc",
    doi = "10.1103/PhysRevD.103.104036",
    journal = "Phys. Rev. D",
    volume = "103",
    number = "10",
    pages = "104036",
    year = "2021"
}

@article{Atamurotov:2013dpa,
    author = "Atamurotov, Farruh and Abdujabbarov, Ahmadjon and Ahmedov, Bobomurat",
    title = "{Shadow of rotating Ho{\v{r}}ava-Lifshitz black hole}",
    doi = "10.1007/s10509-013-1548-5",
    journal = "Astrophys. Space Sci.",
    volume = "348",
    pages = "179--188",
    year = "2013"
}

@article{Liu:2025lwj,
    author = "Liu, Wentao and Huang, Hongxia and Wu, Di and Wang, Jieci",
    title = "{Lorentz violation signatures in the low-energy sector of Ho{\v{r}}ava gravity from black hole shadow observations}",
    eprint = "2506.13504",
    archivePrefix = "arXiv",
    primaryClass = "gr-qc",
    doi = "10.1016/j.physletb.2025.139812",
    journal = "Phys. Lett. B",
    volume = "868",
    pages = "139812",
    year = "2025"
}

@article{Zhao:2025nwi,
    author = "Zhao, Meng-Die and Wang, Yu-Yan and He, Ke-Jian and Li, Guo-Ping",
    title = "{Probing observable features of Lorentz violation in low-energy Ho{\v{r}}ava gravity with accretion disk images of black hole}",
    eprint = "2511.23117",
    archivePrefix = "arXiv",
    primaryClass = "gr-qc",
    doi = "10.1140/epjc/s10052-026-15545-y",
    journal = "Eur. Phys. J. C",
    volume = "86",
    number = "4",
    pages = "329",
    year = "2026"
}

@article{Hu:2020usx,
    author = "Hu, Zezhou and Zhong, Zhen and Li, Peng-Cheng and Guo, Minyong and Chen, Bin",
    title = "{QED effect on a black hole shadow}",
    eprint = "2012.07022",
    archivePrefix = "arXiv",
    primaryClass = "gr-qc",
    doi = "10.1103/PhysRevD.103.044057",
    journal = "Phys. Rev. D",
    volume = "103",
    number = "4",
    pages = "044057",
    year = "2021"
}

@article{Wang:2025ihg,
    author = "Wang, Xinyu and Zhao, Zhixing and Zeng, Xiao-Xiong and Wang, Xin-Yang",
    title = "{Revisiting the shadow of Johannsen-Psaltis black holes}",
    eprint = "2501.08287",
    archivePrefix = "arXiv",
    primaryClass = "gr-qc",
    doi = "10.1103/PhysRevD.111.084054",
    journal = "Phys. Rev. D",
    volume = "111",
    number = "8",
    pages = "084054",
    year = "2025"
}

@article{Peng:2020wun,
    author = "Peng, Jun and Guo, Minyong and Feng, Xing-Hui",
    title = "{Influence of quantum correction on black hole shadows, photon rings, and lensing rings}",
    eprint = "2008.00657",
    archivePrefix = "arXiv",
    primaryClass = "gr-qc",
    doi = "10.1088/1674-1137/ac06bb",
    journal = "Chin. Phys. C",
    volume = "45",
    number = "8",
    pages = "085103",
    year = "2021"
}

@article{Hou:2021okc,
    author = "Hou, Yehui and Guo, Minyong and Chen, Bin",
    title = "{Revisiting the shadow of braneworld black holes}",
    eprint = "2103.04369",
    archivePrefix = "arXiv",
    primaryClass = "gr-qc",
    doi = "10.1103/PhysRevD.104.024001",
    journal = "Phys. Rev. D",
    volume = "104",
    number = "2",
    pages = "024001",
    year = "2021"
}

@article{Zhang:2023bzv,
    author = "Zhang, Zhenyu and Hou, Yehui and Guo, Minyong",
    title = "{Observational signatures of rotating black holes in the semiclassical gravity with trace anomaly*}",
    eprint = "2305.14924",
    archivePrefix = "arXiv",
    primaryClass = "gr-qc",
    doi = "10.1088/1674-1137/ad432b",
    journal = "Chin. Phys. C",
    volume = "48",
    number = "8",
    pages = "085106",
    year = "2024"
}

@article{Guo:2020zmf,
    author = "Guo, Minyong and Li, Peng-Cheng",
    title = "{Innermost stable circular orbit and shadow of the $4D$ Einstein{\textendash}Gauss{\textendash}Bonnet black hole}",
    eprint = "2003.02523",
    archivePrefix = "arXiv",
    primaryClass = "gr-qc",
    doi = "10.1140/epjc/s10052-020-8164-7",
    journal = "Eur. Phys. J. C",
    volume = "80",
    number = "6",
    pages = "588",
    year = "2020"
}

@article{Guo:2018kis,
    author = "Guo, Minyong and Obers, Niels A. and Yan, Haopeng",
    title = "{Observational signatures of near-extremal Kerr-like black holes in a modified gravity theory at the Event Horizon Telescope}",
    eprint = "1806.05249",
    archivePrefix = "arXiv",
    primaryClass = "gr-qc",
    doi = "10.1103/PhysRevD.98.084063",
    journal = "Phys. Rev. D",
    volume = "98",
    number = "8",
    pages = "084063",
    year = "2018"
}
		
\end{document}